\documentclass[%
 reprint,
 amsmath,amssymb,
 prb,
]{revtex4-1}

\usepackage{graphicx}
\usepackage{dcolumn}
\usepackage{bm}

\renewcommand\vec[1]{{\bf #1}}

\begin{document}


\title{Band offsets, wells and barriers at nanoscale semiconductor heterojunctions}

\author{Yann-Michel Niquet}
\affiliation{Laboratoire de simulation atomistique (L\_Sim), SP2M, UMR-E CEA/UJF-Grenoble 1, INAC, Grenoble, F-38054, France}
\email{yann-michel.niquet@cea.fr}

\author{Christophe Delerue}
\affiliation{IEMN - Dept. ISEN, 41 boulevard Vauban, 59046 Lille Cedex France}
\email{christophe.delerue@isen.fr}

\date{\today}

\begin{abstract}
Epitaxially-grown semiconductor heterostructures give the possibility to tailor the potential landscape for the carriers in a very controlled way. In planar lattice-matched heterostructures, the potential has indeed a very simple and easily predictable behavior: it is constant everywhere except at the interfaces where there is a step (discontinuity) which only depends on the composition of the semiconductors in contact. In this paper, we show that this universally accepted picture can be invalid in nanoscale heterostructures (e.g., quantum dots, rods, nanowires) which can be presently fabricated in a large variety of forms. Self-consistent tight-binding calculations applied to systems containing up to 75\,000 atoms indeed demonstrate that the potential may have a more complex behavior in axial hetero-nanostructures: The band edges can show significant variations far from the interfaces if the nanostructures are not capped with a homogeneous shell. These results suggest new strategies to engineer the electronic properties of nanoscale objects, e.g. for sensors and photovoltaics.
\end{abstract}

\pacs{Valid PACS appear here}
\maketitle

\section{Introduction}

For the last four decades, semiconductor heterostructures have been at the heart of major scientific discoveries (e.g., quantum Hall effect) and technological innovations (e.g., resonant tunneling diodes, solid-state lasers, high-frequency electronic devices). The epitaxial growth of successive semiconductor layers with different band gaps indeed introduces energy barriers and wells for the carriers, which control their optical and transport properties.\cite{Bastard88,Burt92,Vurgaftman01} In planar lattice-matched heterostructures, the potential experienced by a carrier in each layer is actually the same as in the corresponding bulk material except for a rigid shift due to the presence of a two-dimensional (2D) dipole layer at the interface.

Recently, low dimensional nanostructures containing multiple semiconductor compounds have been synthesized, bringing forth a new generation of materials with unique electronic and optical properties. These include 0D (e.g., core/shell nanocrystals, \cite{Peng97,Dabboussi97,Mekis03,Mahler08} nanorods, \cite{Talapin03,Sadtler09,Borys10} tetrapods, \cite{Chin07,Talapin07} dumbbells, \cite{Kudera05} stars \cite{Lee02}) and 1D structures (e.g., nanowire super-lattices, \cite{Bjork02,Gudiksen02,Dick10,Carnevale11} core/shell nanowires \cite{Lauhon02,Noborisaka05,Skold05}). The usual paradigm for the electronic structure of these complex objects is that the potential profile in a nanoscale heterojunction is exactly the same as in the planar case. \cite{Note_relaxations} This assumption underlies all non self-consistent, semi-empirical descriptions of semiconductor nanostructures (effective-mass, ${\bf k}.{\bf p}$, \cite{Haus93,Garcia08,Pistol09,Lue09} tight-binding, \cite{Perez03,Diaz06,Niquet08} pseudopotentials \cite{Schrier06,Zhang10}), widely used for their relative simplicity. But the application of the concepts of band offsets, potential wells and barriers to nanostructures is questionable. \cite{Leonard00} In this paper, we actually show that nanoscale heterostructures may require more elaborate treatments depending on their dimensionality, their symmetry, and their surface passivation. We demonstrate that the potential in these systems presents the same discontinuity at the interface as in the planar case but does not necessarily behave as a simple step function. The variations of the potential away from the interface may be substantial and therefore strongly influence the electronic states. This provides new opportunities to engineer the properties of nanostructures.

The paper is organized as follows: we describe the methodology in section \ref{sectionMethodology}, then set out the main conclusions from an analysis of various examples with different dimensionalities (nanocrystals, nanowires, ...) in section \ref{sectionResults}. We discuss the underlying physics and present a more quantitative theory in section \ref{sectionDiscussion}. We finally discuss other nanostructures with mixed dimensionalities as well as possible applications in section \ref{sectionOthers}.

\section{Methodology}
\label{sectionMethodology}

In the following, we consider GaAs/AlAs as a prototypical system. The relevant physics is, indeed, most easily highlighted in binary, lattice-matched materials, where the band edges do not show additional variations due to inhomogeneous strains and alloy disorder. The conclusions of this work however apply to heterostructures of any semiconductor compounds or alloys. The band offset problem is actually challenging because only a self-consistent electronic structure method can allow for charge transfers between materials -- yet any realistic nanoscale heterojunction contains at least thousands of atoms and is beyond the reach of {\it ab initio} approaches such as density functional theory. It has been shown previously that self-consistent tight-binding accurately describes electrostatic and screening effects in semiconductors. \cite{Delerue97,Delerue03} Here we use this approach, that we have considerably optimized thanks to efficient algorithms which enable the treatment of nanostructures with more than 75\,000 atoms.

\subsection{The self-consistent tight-binding method}

The electronic states of the nanostructures are described with a $sp^3d^5s^*$ tight-binding model. The hamiltonian matrix is written $H=H_{0}+V_{0}+V^{\rm sc}$ where $H_0$ is the bare hamiltonian calculated using the tight-binding parameters of Ref.~\onlinecite{Jancu98} which provide excellent band structures for GaAs and AlAs. Since in this parametrization the energy of the valence-band (VB) edge of each semiconductor is arbitrarily set to zero, we apply a rigid shift $V_0$ to the AlAs atomic energies in order to define the relative position of the bands of the two materials prior to self-consistency. $V_0$ is adjusted to achieve a VB offset of 447 meV for the GaAs/AlAs (110) interface, in agreement with the {\it ab initio} calculations of Ref.~\onlinecite{Bylander87}. But it is important to point out that the choice of $V_0$ (within a reasonable range) has a little influence on the results presented in this work \cite{note_offset_shift} for reasons discussed in Refs. \onlinecite{Flores87,Tersoff84}. $V^{\rm sc}$ is the potential induced by the variations of the atomic charges with respect to the bulk references. These charges are computed from the Green function of the system\cite{Brandbyge02} using an efficient ``knitting'' algorithm\cite{Kazymyrenko08} (see details in the Appendix). The potential is calculated iteratively until self-consistency is achieved. In this paper, we discuss the variations of the total potential $V_{0}+V^{\rm sc}$ which, with the above conventions, can be intepreted as the local VB edge in the nanostructure (or equivalently as the confinement potential for the holes).

\subsection{Application to GaAs/AlAs super-lattices}

\begin{figure}
\includegraphics[width = .90\columnwidth]{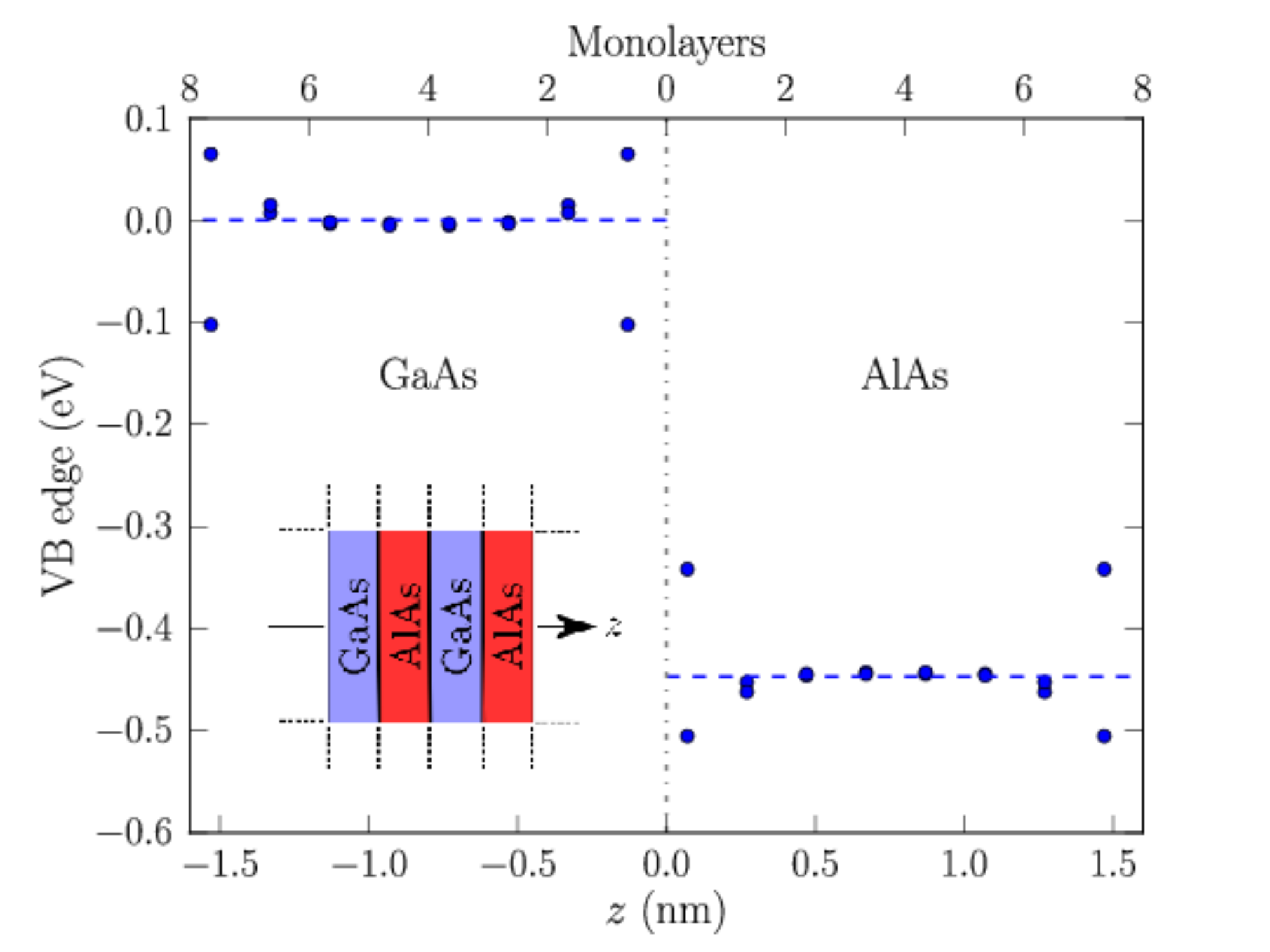}
\caption{\label{fig_superlattice} VB edge in a (GaAs)$_8$(AlAs)$_8$ (110) super-lattice. The average potential in the GaAs layer has been shifted to zero, so that the horizontal dashed lines directly indicate the VB offset.}
\end{figure}

As an illustration, the VB edge in a (GaAs)$_8$(AlAs)$_8$ (110) super-lattice is plotted in Fig.~\ref{fig_superlattice}. It shows the expected behavior of a periodic step function with sharp variations limited to two atomic planes on each side of the interfaces. Elsewhere, the potential is basically constant. Very similar results are obtained for systems with larger periods and for (111) and (001) super-lattices.

\subsection{Surface passivation of nanostructures}
\label{subsectionpassivation}

In the following, we consider finite-size heterostructures such as nanocrystals and nanowires. There the band edges are also controlled by the surface dipoles which depend on the passivation. Recent experiments have actually shown that the nature of the ligands determines the ionization potential of nanocrystals. \cite{Soreni08,Aldakov06} The work function of semiconductor surfaces can indeed be modified (up to 1 eV in the case of GaAs \cite{Bastide97}) by grafting polar molecules, \cite{Ashkenasy02} and, more generally, by various chemical or technological treatments. In our calculations, we emulate these treatments by passivating the dangling bonds at the surface of the nanostructures with hydrogen atoms. The tight-binding parameters of these atoms determine the charge transfers with the semiconductor, hence the surface dipole layer which controls the absolute value of the inner potential. We can therefore tune the ionization potential and mimick the effect of different ligands (or capping materials) by changing the parameters of these ``pseudo''-hydrogen atoms (given in the Appendix). In the next section, we discuss different passivations when appropriate.

\section{Results}
\label{sectionResults}

In this section, we highlight the main conclusions of this work on various examples with different dimensionalities: 0D core/shell quantum dots (paragraph \ref{subsectionDots}) and 1D nanowires (paragraph \ref{subsectionWires}). We briefly dicuss the physics behind these examples, and give a more quantitative and complete analysis in section \ref{sectionDiscussion}.

\subsection{Core/shell quantum dots}
\label{subsectionDots}

\begin{figure}
\includegraphics[width = .90\columnwidth]{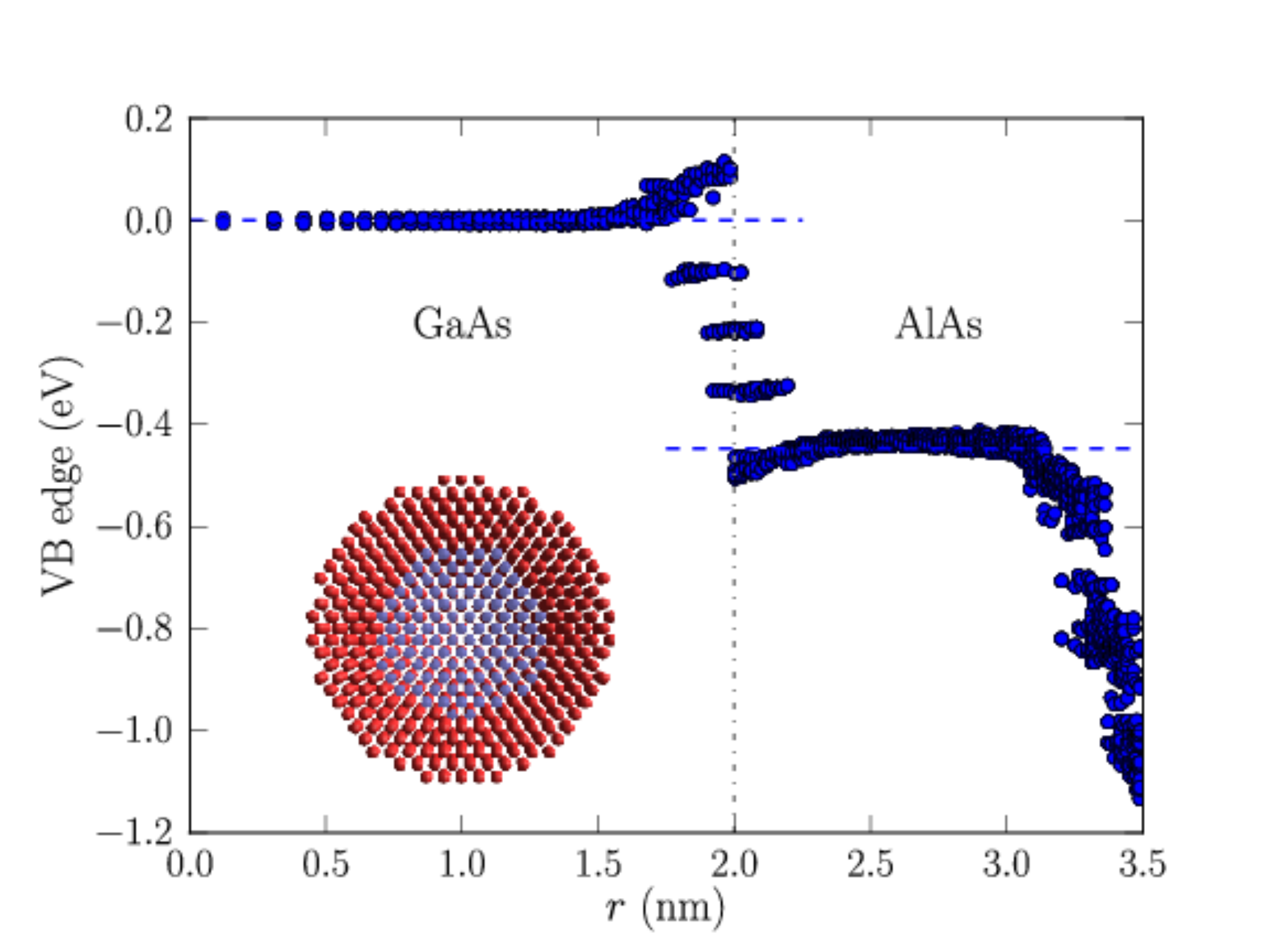}
\caption{\label{fig_cs_sphere} VB edge in a spherical GaAs/AlAs core/shell quantum dot as a function of the radial position of the atoms (core diameter $d_c=4$ nm; shell diameter $d_s=7$ nm). The horizontal dashed lines indicate the VB offset in the 2D GaAs/AlAs super-lattice.}
\end{figure}

Figure~\ref{fig_cs_sphere} presents a typical result for GaAs/AlAs core/shell quantum dots. The rapid variation of the potential at the surface is due to the charge transfers with pseudo-hydrogen atoms as discussed above. The VB edge is constant in the core and in the shell, respectively, and the VB discontinuity is the same as for the planar interface. Therefore the traditional model of square wells and barriers is justified in spherical core/shell quantum dots, as well as the use of the band offset derived from the 2D case (for core diameter $\gtrsim 1.5$ nm and shell thickness $\gtrsim 0.75$ nm). In addition, we have verified that the band offset at the interface does not depend on the dipole layer at the surface, thus on the nature of the ligands.

\subsection{Nanowires/rods with axial heterostructures}
\label{subsectionWires}

\subsubsection{Nanowires/rods without external shell}

\begin{figure}
\includegraphics[width = .90\columnwidth]{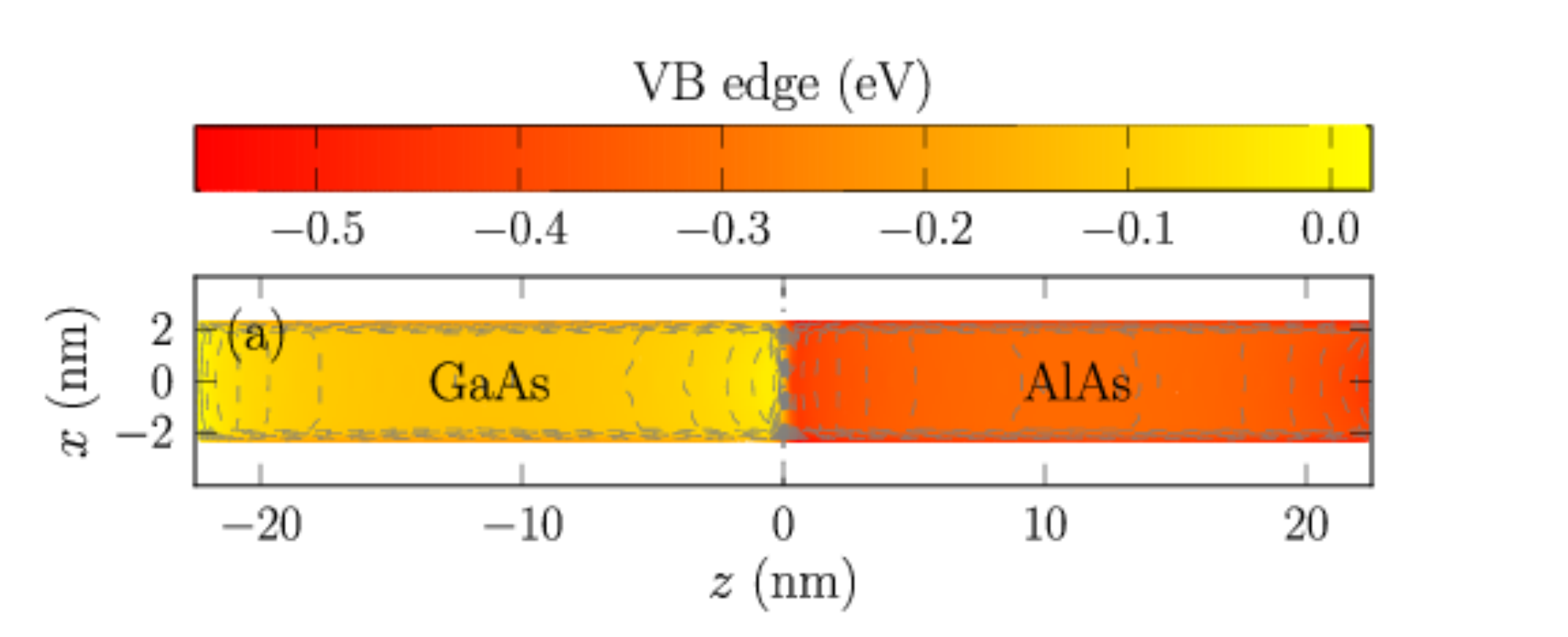}
\includegraphics[width = .90\columnwidth]{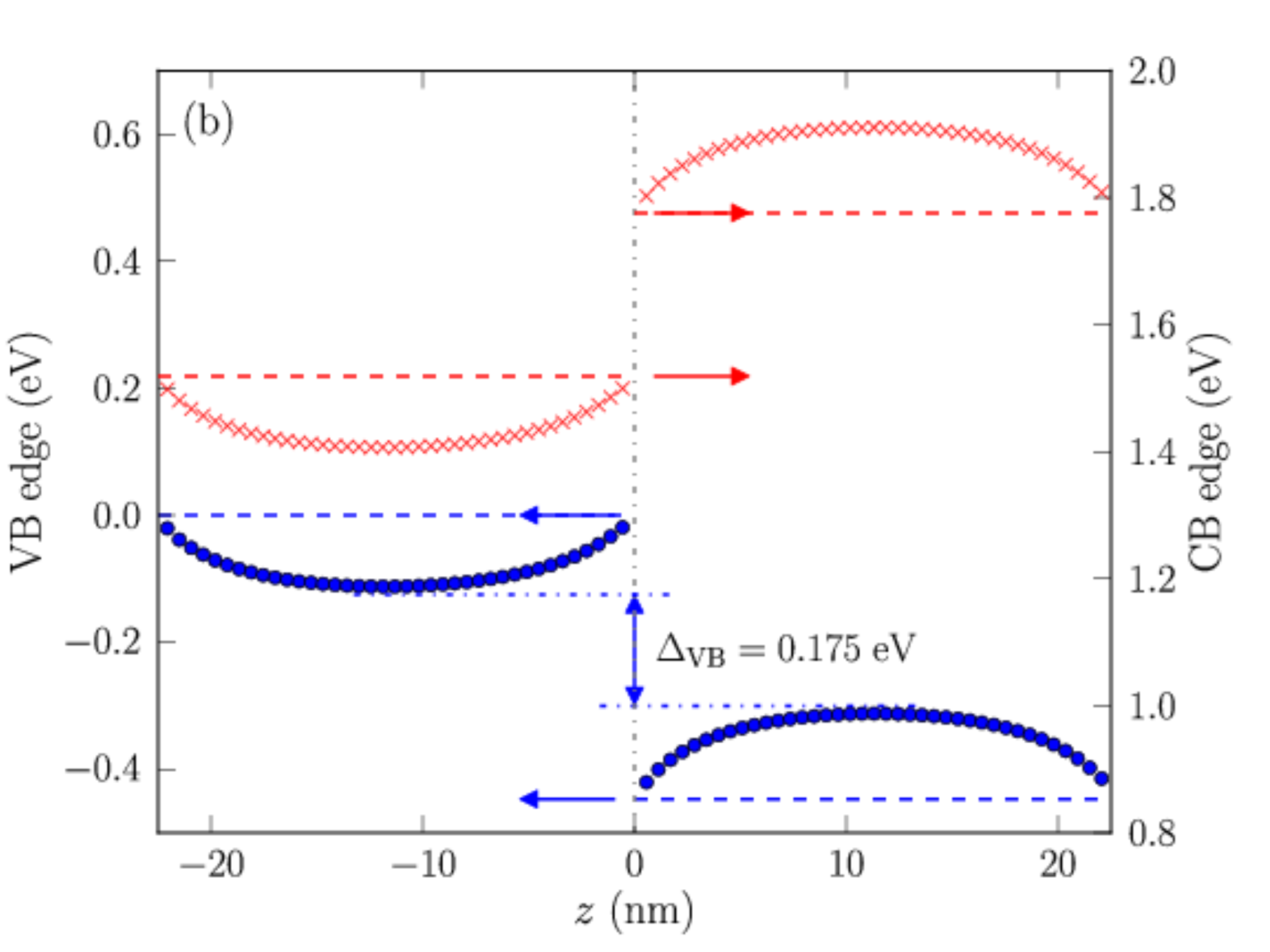}
\includegraphics[width = .90\columnwidth]{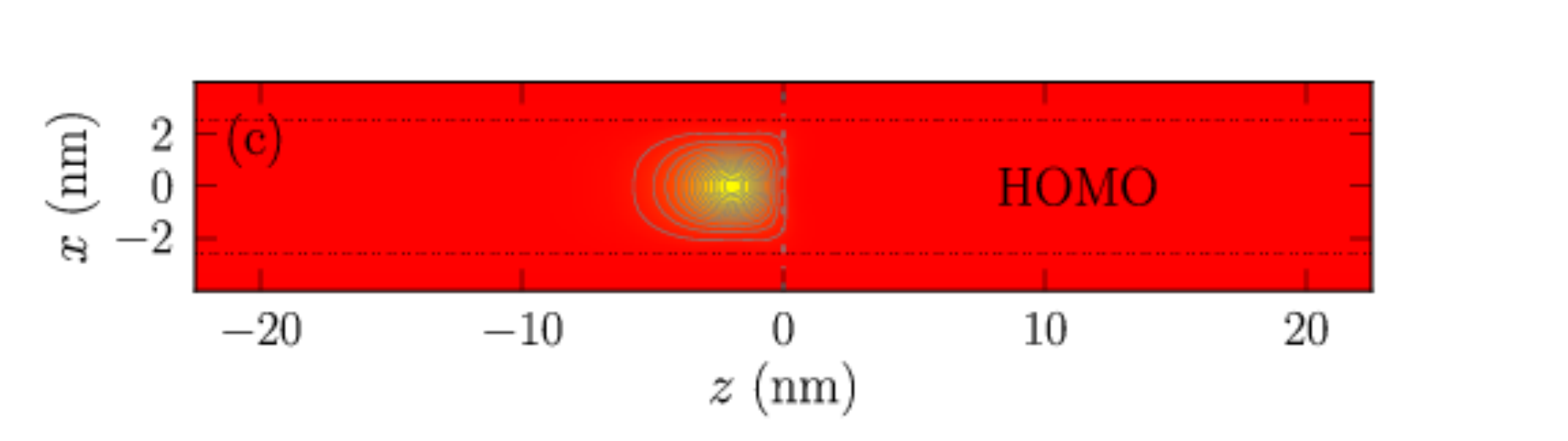}
\includegraphics[width = .90\columnwidth]{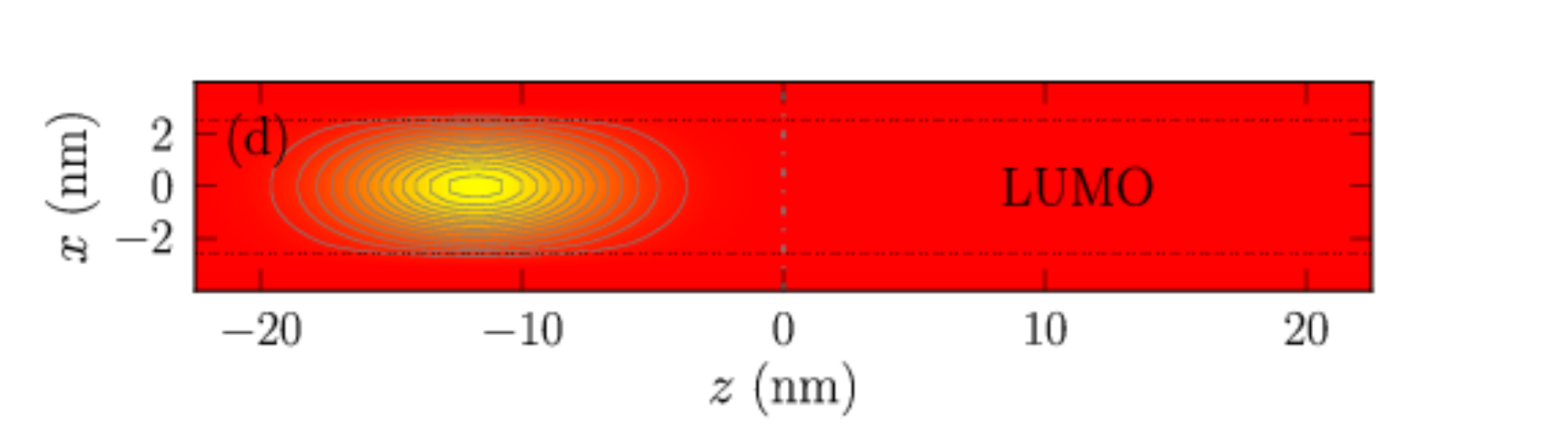}
\caption{\label{fig_wire} a) 2D plot of the VB edge in a section of a GaAs/AlAs nanowire super-lattice with cylindrical shape (diameter $d=5$ nm, length of the GaAs and AlAs segments $l=22.5$ nm). b) VB edge (blue dots) and CB edge (red crosses) along the axis of the nanowire super-lattice. The horizontal dashed lines indicate the VB and CB offsets in the 2D GaAs/AlAs super-lattice. The difference $\Delta_{\rm VB}$ between the VB edges in GaAs and AlAs drops to 0.175 eV in between the interfaces, and would tend to 0.1 eV when $l\to\infty$, the difference between the VB edges of the pristine GaAs and AlAs nanowires. c) and d) 2D plots of the envelopes of the highest occupied and of the lowest unoccupied states, respectively.}
\end{figure}

\begin{figure}
\includegraphics[width = .90\columnwidth]{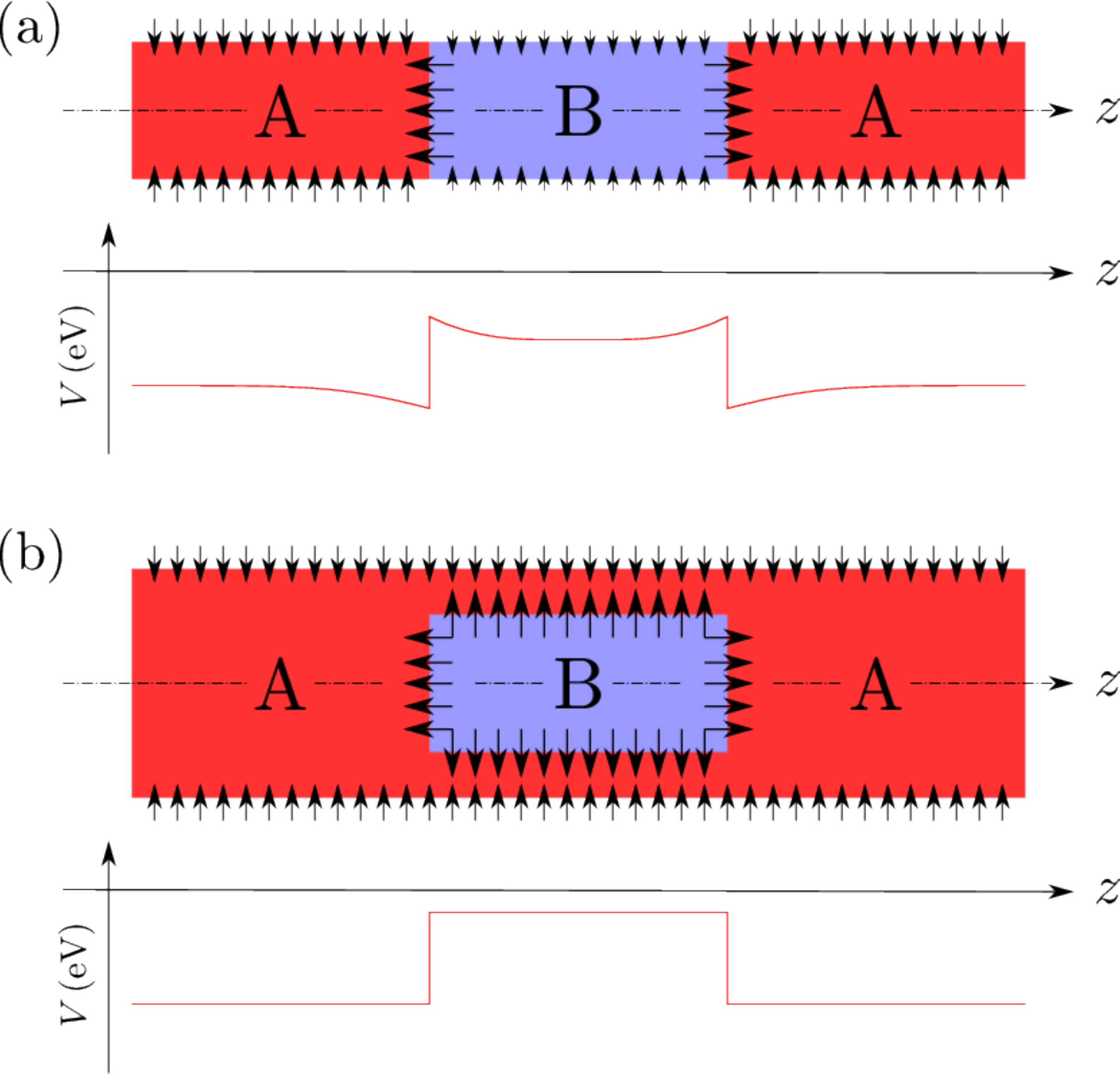}
\caption{\label{fig_dipoles} Schematic representation of the dipole layers at the surfaces and interfaces of an A/B nanowire super-lattice with (a) or without (b) a shell. The contribution of these dipoles to the band offset is also plotted.}
\end{figure}

Similar results are obtained for core/shell GaAs/AlAs nanowires with cylindrical or facetted geometries. The situation is, however, very different in 1D (nanowires) or nearly 1D (e.g., nanorods) systems embedding axial heterostructures. Figure~\ref{fig_wire} presents typical results for a nanowire super-lattice made of successive segments of GaAs and AlAs. The VB discontinuity right at the interface is once again the same as in the planar case but the VB edge is not constant along the nanowire axis. Indeed, as shown by L\'eonard and Tersoff in carbon-nanotube Schottky diodes, \cite{Leonard00} the dipole layer at the interface shifts the potential only over distances comparable with its lateral dimension (the diameter of the nanowire). Far from this interface, the potential is controlled by the surface dipoles (see Fig.~\ref{fig_dipoles}), so that the VB edge tends to the value for the corresponding pristine nanowire: for the passivation considered here, the VB edge of the pristine AlAs nanowire is only $\simeq0.1$ eV lower than the VB edge of the pristine GaAs nanowire (in other words, the difference between the VB edges of the pristine GaAs and AlAs nanowires is $\simeq0.35$ eV smaller than the planar band offset). The important variation of the VB and CB edges displayed in Fig.~\ref{fig_wire}b has a significant effect on the electronic structure of the nanowires. Figure~\ref{fig_wire}c shows that the highest occupied level is strongly localized at the GaAs side of the interface while the lowest unoccupied state remains centered in the GaAs segment. As an interesting side effect, a barrier is raised at the interface for the holes in the AlAs segment.

\begin{figure}
\includegraphics[width = .90\columnwidth]{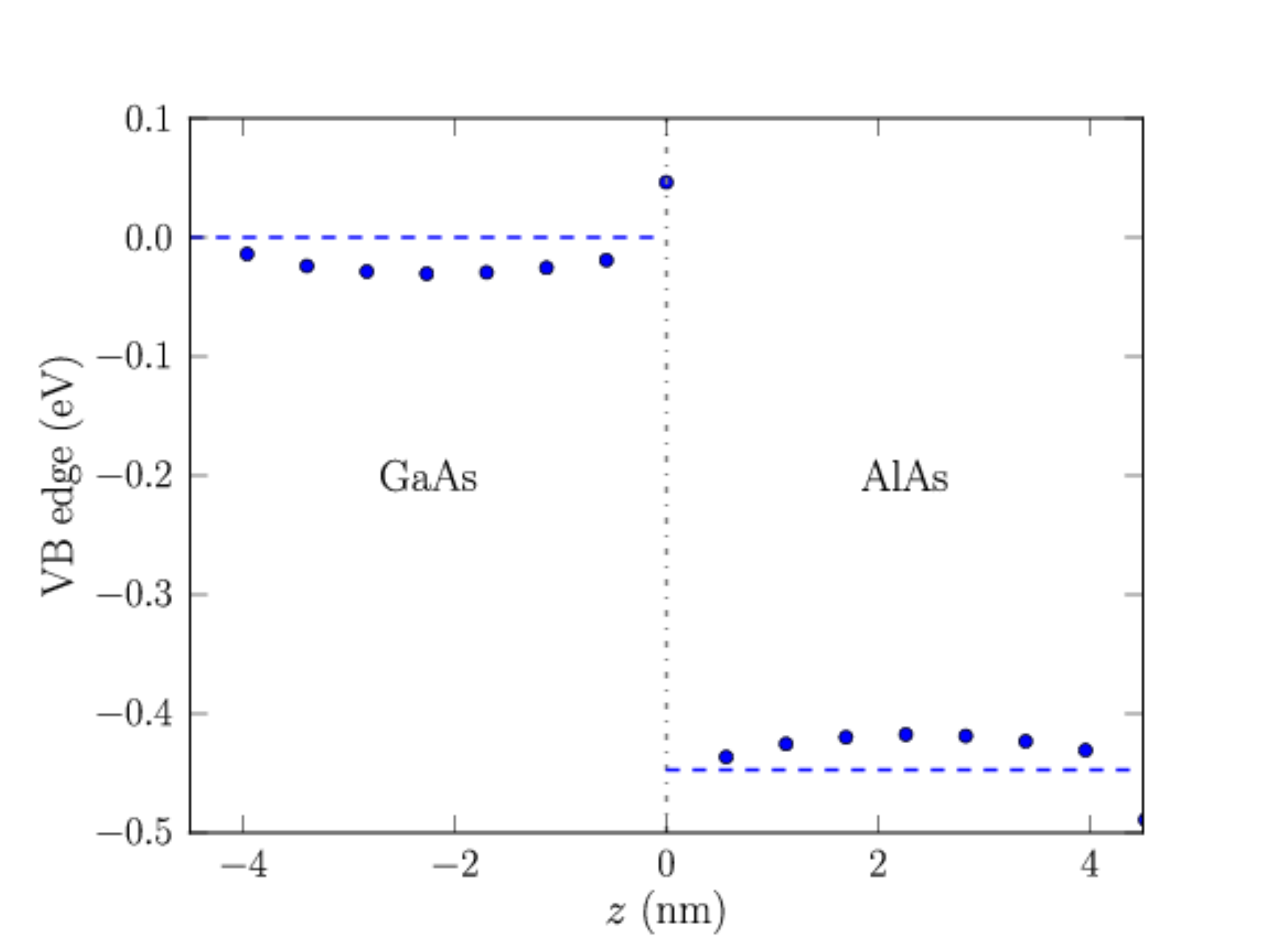}
\caption{\label{fig_wire2} VB edge along the axis of a GaAs/AlAs nanowire super-lattice with cylindrical shape and short period (diameter $d=5$ nm, length of the GaAs and AlAs segments $l=4.5$ nm). The tight-binding parameters are the same as for Fig.~\ref{fig_wire} (long period counterpart).}
\end{figure}

\begin{figure}
\includegraphics[width = .90\columnwidth]{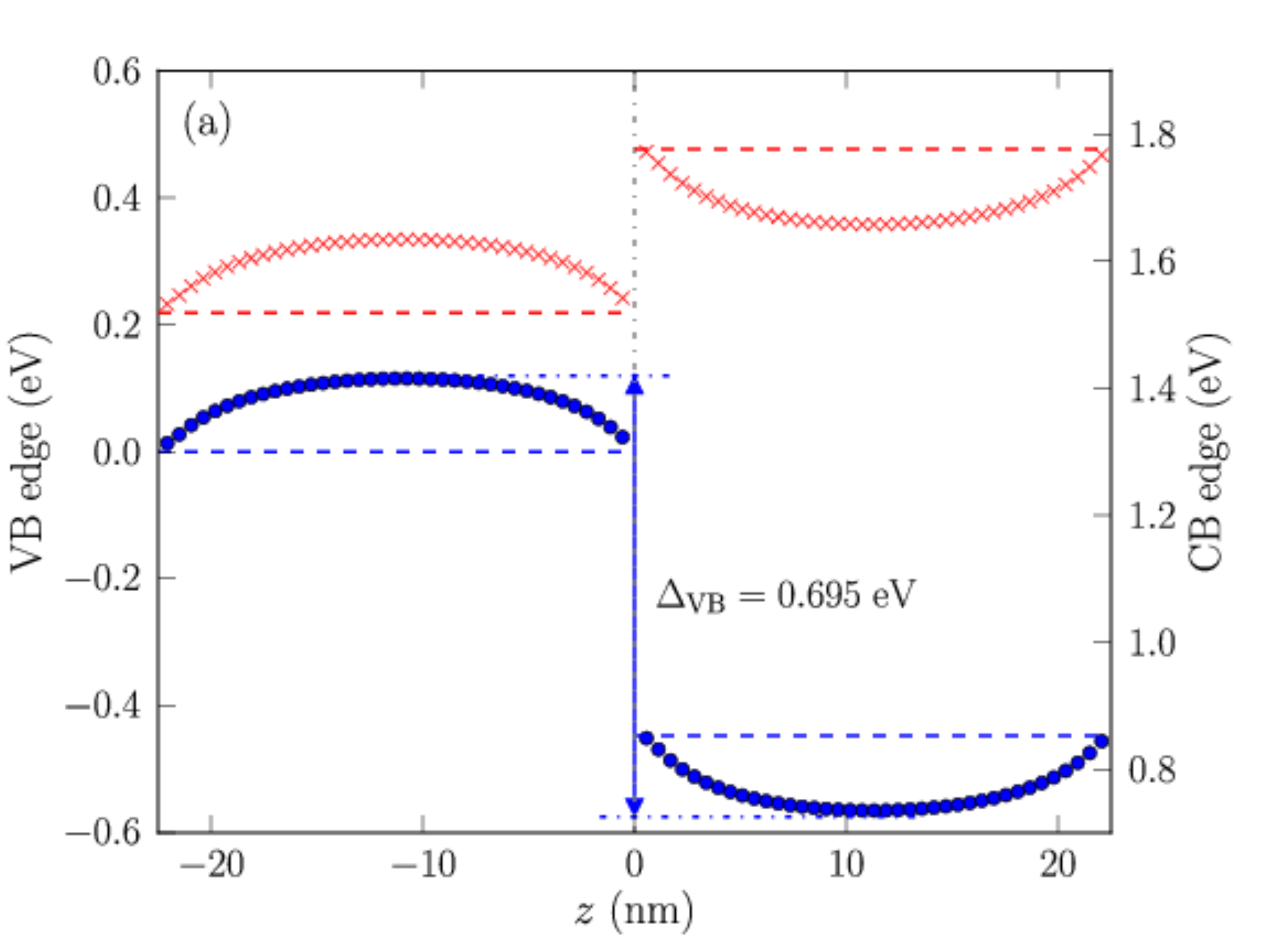}
\includegraphics[width = .90\columnwidth]{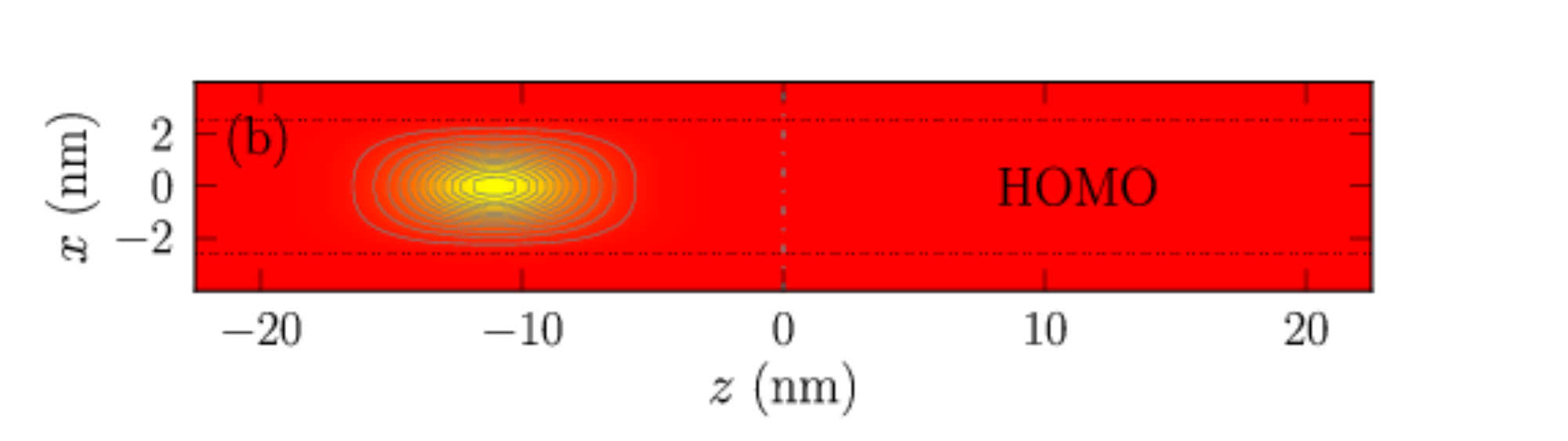}
\includegraphics[width = .90\columnwidth]{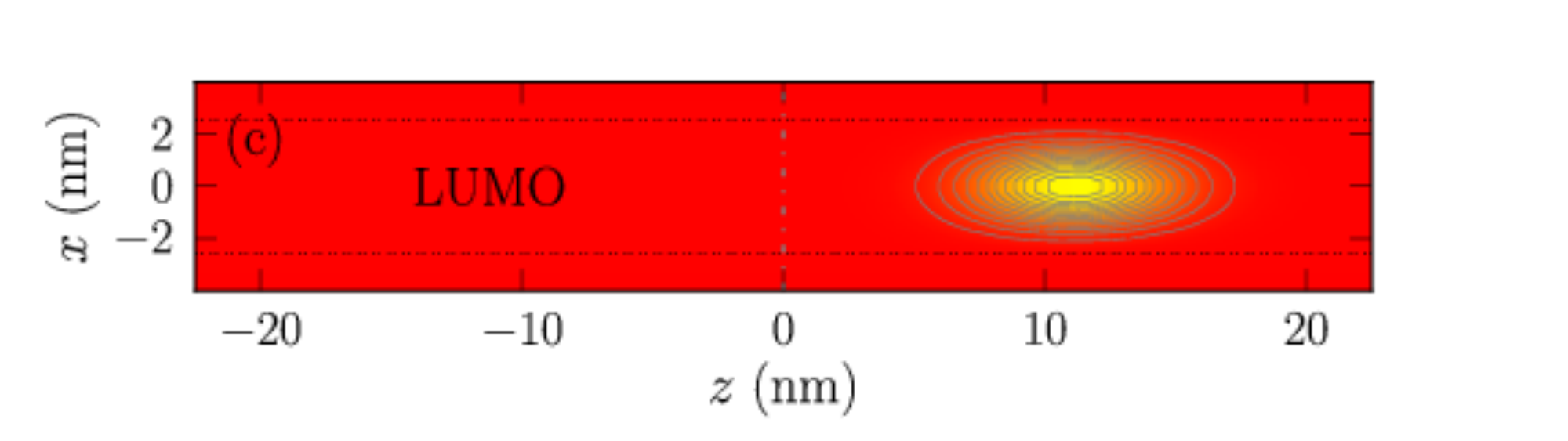}
\caption{\label{fig_wire_inv} Same as Fig.~\ref{fig_wire} but after modification of the pseudo-hydrogen atoms at the surface (see text). a) VB edge (blue dots) and CB edge (red crosses). The difference $\Delta_{\rm VB}$ between the VB edges in GaAs and AlAs reaches 0.695 eV in between the interfaces, and would tend to 0.8 eV when $l\to\infty$, the difference between the VB edges of the pristine GaAs and AlAs nanowires. b) and c) 2D plots of the envelopes of the highest occupied and of the lowest unoccupied states, respectively.}
\end{figure}

The effects highlighted in Fig.~\ref{fig_wire} are mostly visible when the period of the super-lattice is larger than the diameter of the nanowire (see Fig.~\ref{fig_wire2} for a comparison with a short period super-lattice). Indeed, in long nanowire segments, the band edge profiles are determined not only by the discontinuity at the heterojunctions but also by the surface passivation (e.g. ligands) which controls the absolute value of the inner potential far from the interfaces. The influence of the surface termination is further demonstrated in Fig.~\ref{fig_wire_inv}. In that case, we have used a different set of parameters for the pseudo-hydrogen atoms, such that the VB edge of the pristine GaAs nanowire is now $\simeq0.8$ eV higher than the VB edge of the pristine AlAs nanowire (i.e. the difference between the VB edges is $\simeq0.35$ eV larger than the planar band offset, opposite to the previous case). The variation of the VB edge in the super-lattice is then inverted with respect to Fig.~\ref{fig_wire}. Interestingly, the lowest unoccupied state is now located in the AlAs segment, even though the CB edge of GaAs remains below the CB edge of AlAs. Indeed, quantum confinement, which is stronger in GaAs due to its small effective mass, raises the electronic states of the GaAs segment above those of the AlAs segment.

\subsubsection{Nanowires/rods with an external shell}

\begin{figure}
\includegraphics[width = .90\columnwidth]{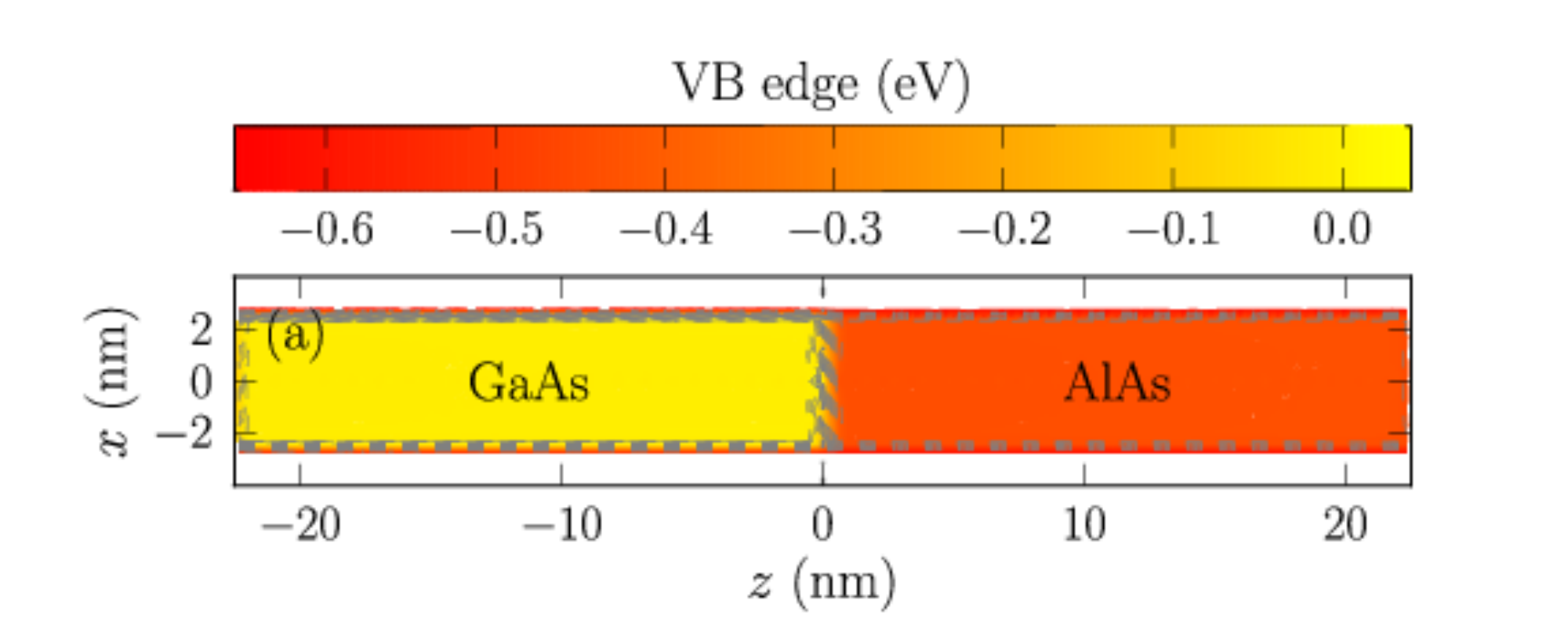}
\includegraphics[width = .90\columnwidth]{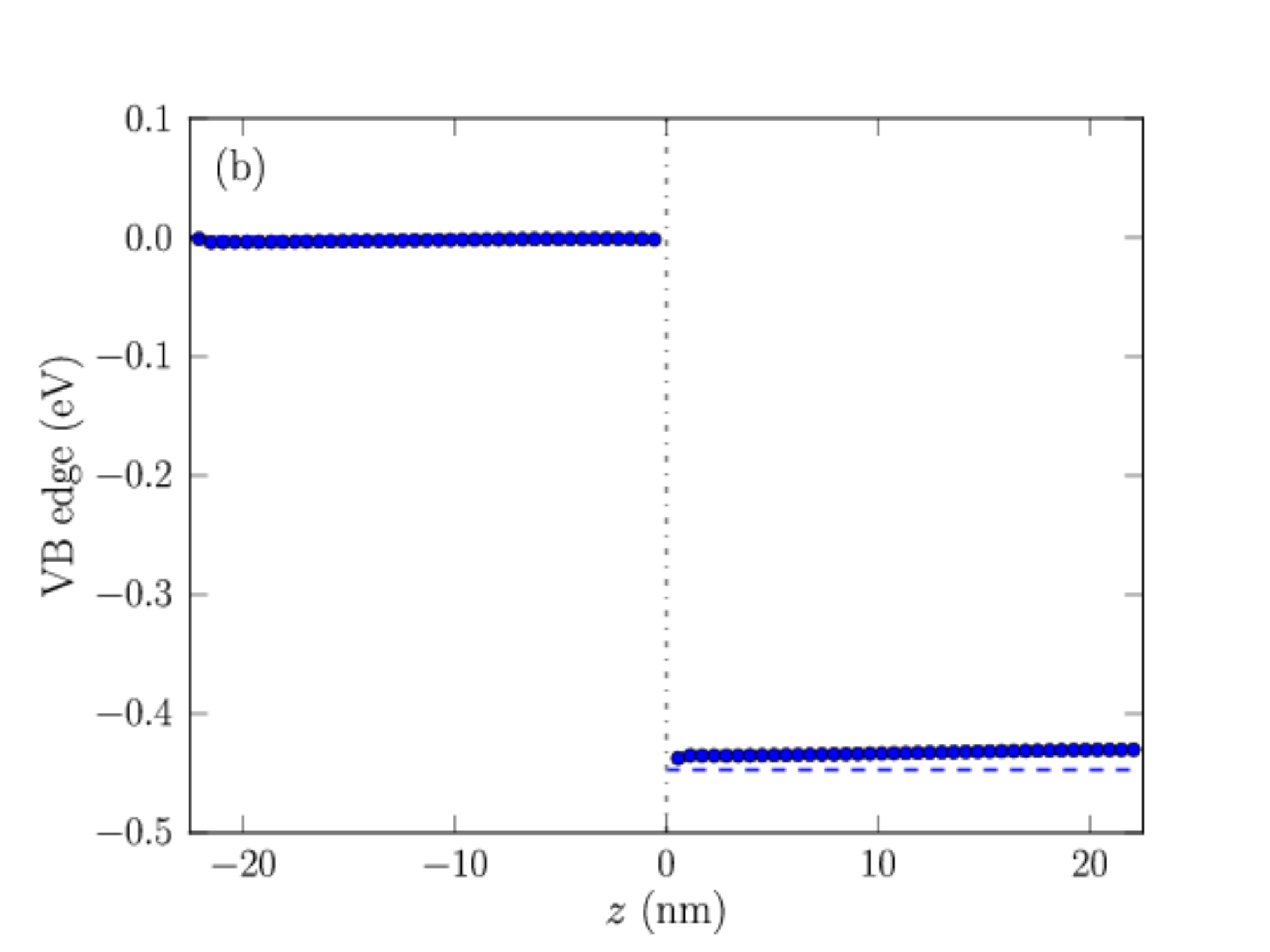}
\caption{\label{fig_cswire} a) 2D plot of the VB edge in a section of a GaAs/AlAs nanowire super-lattice surrounded by an AlAs shell (core diameter $d_c=5$ nm, shell diameter $d_s=6$ nm; length of each GaAs or AlAs segment $l=22.5$ nm). b) VB edge along the axis of the nanostructure. The horizontal dashed lines indicate the VB offset in the 2D GaAs/AlAs super-lattice.}
\end{figure}

The importance of the surface termination is also illustrated in Fig.~\ref{fig_cswire} which represents a nanowire super-lattice now surrounded by an additional shell of AlAs. Then a simple behavior is recovered: the VB edge is constant everywhere except at the GaAs/AlAs interfaces where there is the expected discontinuity. The uniform dipole layer at the surface and interfaces (see Fig.~\ref{fig_dipoles}) rigidly shifts the potential in the nanowire but do not influence its form.

In the next section, we give a more quantitative and comprehensive picture of the above results, and clarify in particular the role of the shell in the band offset problem.

\section{Discussion}
\label{sectionDiscussion}

We can draw a few rules about the behavior of the potential in nanoscale heterostructures from the different systems we have considered. In principle, the potential is determined in a complex manner by the geometry and composition of the nanostructure, and by its surface termination. For example, the Madelung potential is not the same in low-dimensional systems as in bulk. Charge transfers also occur between the different materials in the nanostructure. However, we observe that the self-consistent potential $V_{\rm sc}$ always has a relatively simple behavior, which can be interpreted as the resultant of dipole layers located at the surfaces and interfaces. Also, the band edge discontinuity right at the interface is remarkably robust and remains very close to the 2D band offset down to extreme nanostructures with characteristic dimensions below $\simeq2$ nm. We discuss in more detail the physics of axial 1D heterostructures in paragraph \ref{subsectionDisc1D}, then clarify the role of the shell in paragraph \ref{subsectionDisccoreshell}. We also discuss why the discontinuity at the interface remains so close to the planar limit in paragraph \ref{subsectionCNL}, and the effects of doping and surface defects (disregarded up to now) in paragraph \ref{subsectionDoping}.

\subsection{Band offsets in 1D axial nanostructures}
\label{subsectionDisc1D}

The situation in axial 1D heterostructures is schematically depicted in Fig.~\ref{fig_dipoles}. In the case of a A/B nanowire super-lattice (where A and B stand for two semiconductor materials), the potential along the axis is determined: {\it i}) by the dipole layer at each A/B interface, which controls the band edge discontinuity; and {\it ii}) by the dipole layer at the surface of the nanowire, which is usually different in A and B segments. The potential created by the interface dipoles is actually short-range (due to the finite cross section of the nanowire) and decays over a few times the radius $R$. Therefore, in a super-lattice with long segments (Figs.~\ref{fig_wire} and \ref{fig_wire_inv}), the surface dipoles prevail over the interface dipoles a few $R$'s away from the heterojunctions, so that the potential far from the A/B interfaces is the same as in the corresponding pristine nanowire and is just determined by the nature of the semiconductor and its capping. On the contrary, a super-lattice with short segments mostly experiences the average distribution of surface dipoles, which can not sustain a significant modulation of the potential deep inside the nanowire (Fig.~\ref{fig_wire2}).

This can be put in a more quantitative way with a simple model neglecting the dielectric mismatch between the nanowire and its environment. The potential $V\equiv V_{\rm sc}$ created by a disk of interface dipoles with density $P$ in a medium with dielectric constant $\varepsilon$ reads, along the symmetry axis $z$:
\begin{equation}
V\left(z\right)=\frac{2\pi P}{\varepsilon}\left[\frac{z-z_i}{\sqrt{R^2+\left(z-z_i\right)^2}}-\frac{z-z_i}{\left|z-z_i\right|}\right]\,,
\label{eqV1}
\end{equation}
where $z_i$ is the position of the disk. It features, as expected, a discontinuity $\Delta V=4\pi P/\varepsilon$ at $z=z_i$, and decays as $V(z)\simeq\pm\pi R^2P/[\varepsilon(z-z_i)^2]$ when $|z-z_i|\gg R$. Likewise, the potential created by a tube of surface dipoles extending from $z=z_i$ to $z=z_j$ reads:
\begin{equation}
V\left(z\right)=\frac{2\pi P}{\varepsilon}\left[\frac{z-z_i}{\sqrt{R^2+\left(z-z_i\right)^2}}-\frac{z-z_j}{\sqrt{R^2+\left(z-z_j\right)^2}}\right]\,.
\label{eqV2}
\end{equation}
In the case of a single A/B interface at $z=0$, the potential therefore behaves on each side of the heterojunction as:
\begin{eqnarray}
V\left(z\right)&\simeq&\frac{4\pi P_{\rm A}}{\varepsilon}-\frac{\pi R^2}{\varepsilon z^2}\left(P_{\rm A}-P_{\rm B}-P_{\rm AB}\right){\rm\ if \ } z\ll -R \nonumber \\
V\left(z\right)&\simeq&\frac{4\pi P_{\rm B}}{\varepsilon}+\frac{\pi R^2}{\varepsilon z^2}\left(P_{\rm A}-P_{\rm B}-P_{\rm AB}\right){\rm\ if \ } z\gg R\,,
\label{eqV3}
\end{eqnarray}
where $P_{\rm A}$ and $P_{\rm B}$ are the dipole densities at the surface of materials A and B, and $P_{\rm AB}$ is the dipole density at the A/B interface. As discussed above, $V(z)$ tends to $4\pi P_{\rm A}/\varepsilon$ in material A, and to $4\pi P_{\rm B}/\varepsilon$ in material B, which are the potentials in pristine A and B nanowires, respectively.

From a practical point of view, the VB and CB edges therefore tend on each side of the interface to their values in pristine A and B nanowires, with a typical $\propto (R/z)^2$ asymptotic behavior. In general, the VB and CB discontinuities at the interface do not match the difference of ionization potentials and affinities so that the band edges do not make a simple step: The difference between the affinities of many semiconductor surfaces can deviate by a few hundreds of meV from the corresponding CB offsets,\cite{Bauer83} and can be enhanced by, e.g., grafting polar molecules which make different dipole distributions on each material.\cite{Ashkenasy02,Bastide97} The band edges in pristine A and B nanowires can be inferred from ionization potential or electron affinity measurements, possibly corrected from confinement effects (the ionization potential is the valence band edge plus the hole confinement energy, while the affinity is the conduction band edge minus the electron confinement energy). The exact band edge profiles away from the axis or in inhomogeneous dielectric environments can be obtained from a numerical solution of Poisson's equation. If the dielectric constants of the two materials are close (which is usually the case), the drop of potential will be about the same on each side of the interface, as evidenced in Figs.~\ref{fig_wire} and \ref{fig_wire_inv}.

Appropriate summation of Eqs.~(\ref{eqV1}) and (\ref{eqV2}) in a single layer with thickness $l$ (or in a super-lattice) also confirms that the potential is only weakly modulated by the surface dipoles when $l\ll R$. The strong modulation of the band edges observed in Figs.~\ref{fig_wire} and \ref{fig_wire_inv} is not, however, limited to nanosize wires, but can actually be evidenced at any $R$ provided the different segments are long enough [since, according to Eq.~(\ref{eqV3}), the potential mostly depends on $z/R$]. In the case of a A/B nanowire super-lattice surrounded by an additional shell of A (Fig.~\ref{fig_dipoles}b), the dipole layer on the outer surface is homogeneous and only induces a rigid shift of the inner potential. Since each B region is completely surrounded by material A, there is just a constant discontinuity at the A/B interface and the potential is basically flat beyond the surfaces and interfaces, as easily verified from Eqs.~(\ref{eqV1}) and (\ref{eqV2}).

\subsection{The role of the shell}
\label{subsectionDisccoreshell}

As discussed above, the band edges make a simple step at the interfaces in core/shell quantum dots and core/shell nanowires (even embedding axial heterostructures). In fact, one can show that the model of square wells and barriers is valid in core-shell nanostructures with {\it arbitrary} geometries when the interfaces are closed shapes with {\it homogeneous} properties (same material or alloy everywhere on each side), as is the case for Figs.~\ref{fig_cs_sphere} and \ref{fig_cswire}. Indeed, the potential inside and outside a {\it closed} surface ${\cal S}$ covered by an uniform distribution of dipoles normal to ${\cal S}$ and with density $P$ is given by (assuming a homogeneous dielectric constant $\varepsilon$):
\begin{equation}
V\left(\vec{r}\right)=\frac{P}{\varepsilon}\int_{\cal S} d^2\vec{r}' \vec{n}\left(\vec{r}'\right)\cdot\frac{\vec{r}'-\vec{r}}{\left|\vec{r}'-\vec{r}\right|^3}=\frac{P}{\varepsilon}\Omega_{\cal S}\left(\vec{r}\right)\,,
\label{eq_PS}
\end{equation}
where $\Omega_{\cal S}(\vec{r})$ is the solid angle subtended by ${\cal S}$ at point $\vec{r}$, i.e. $4\pi$ if $\vec{r}$ is inside the volume delimited by ${\cal S}$ and $0$ if it is outside. Therefore, the potential is constant everywhere except for a discontinuity at the surface. Actually, the argument also holds if the dielectric constant is not homogeneous, but makes a jump $\varepsilon_{\rm in}\to\varepsilon_{\rm out}$ across the surface ${\cal S}$.

This shows that the model of square wells and barriers is valid in core-shell heterostructures, whatever their shape, provided (sufficient conditions) that {\it i}) the surfaces and interfaces are closed shapes, {\it ii}) the dipoles are normal to these surfaces and interfaces, and {\it iii}) their strength is uniform enough. We have verified that fluctuations in the thickness of the shell(s) lead to very small variations (a few meV) of the band edges as long as they remain at least $\simeq5$ \AA\ thick everywhere.

\subsection{Invariance of the band offsets in nanostructures}
\label{subsectionCNL}

In all cases, we have found that the discontinuity right at the interface remains remarkably close to the planar limit (at least for nanocrystals and nanowires with diameters $\gtrsim 2$ nm). This behavior can be explained by the alignment of the so-called charge-neutrality levels on each side of the interface.\cite{Flores87,Tersoff84} The alignment of these levels is mainly governed by local neutrality arguments, and therefore weakly depends on size, dimensionality, and shape of the nanostructure. In our calculations, these arguments also explain the weak dependence of the discontinuity on $V_0$. \cite{note_offset_shift} In contrast, the variations of the band edges beyond the interface depend on the environment, and in particular on the distribution of surface dipoles which compete with the interface dipoles.

\subsection{Effects of defects and doping}
\label{subsectionDoping}

The conclusions drawn in this work also apply to defective nanostructures. Charged surface defects would primarily shift the potential, contributing the imbalance between the electron affinities or ionization energies of, e.g., the different segments of a nanowire heterostructure (especially if the density of surface states is not homogenenous). If the nanostructures are not intrinsic (intentional or non-intentional doping by residual impurities or surface traps), the free carriers might partly screen the variations of the band edges. This effect is well-understood already in planar heterostructures and can be treated with usual approaches (e.g., effective mass Schrodinger-Poisson approximation for the free carriers), using the bare (unscreened) valence band edge profile computed following the lines of this work as input.

\section{Other nanostructures and applications}
\label{sectionOthers}

With the help of the arguments of section \ref{sectionDiscussion}, we can design nanostructures where to expect significant modulations of the band edges for applications to, e.g., photovoltaics or sensors. We discuss an original shell design in paragraph \ref{subsectionShellnoshell}, then nanostructures with mixed 0D/1D dimensionality (dot-rod and dumbbell heterostructures) in paragraph \ref{subsectionMixed}.

\subsection{Nanowires with an inhomogeneous shell}
\label{subsectionShellnoshell}

\begin{figure}
\includegraphics[width = .90\columnwidth]{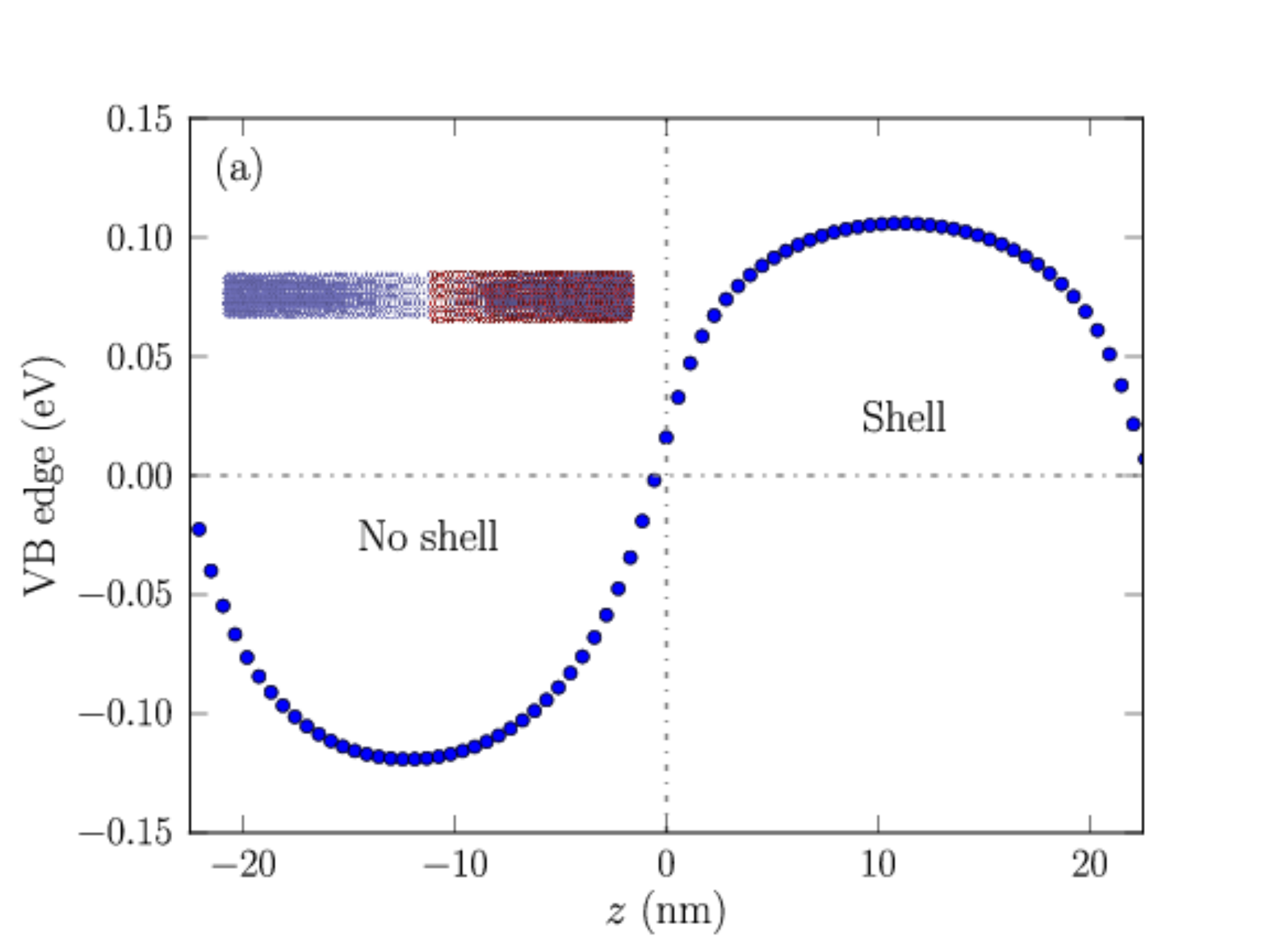}
\includegraphics[width = .90\columnwidth]{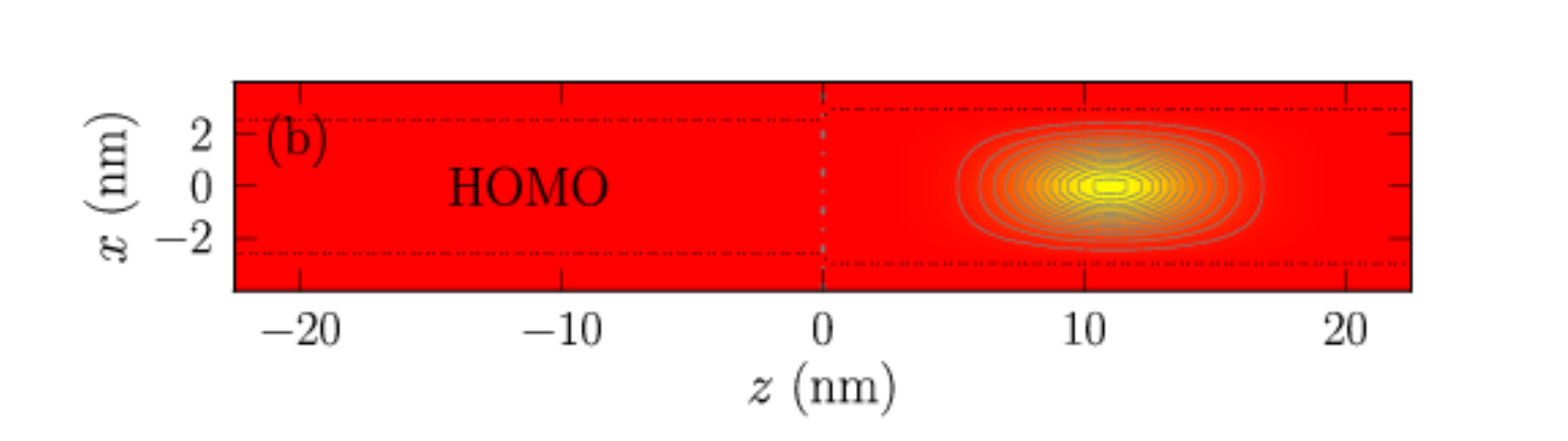}
\includegraphics[width = .90\columnwidth]{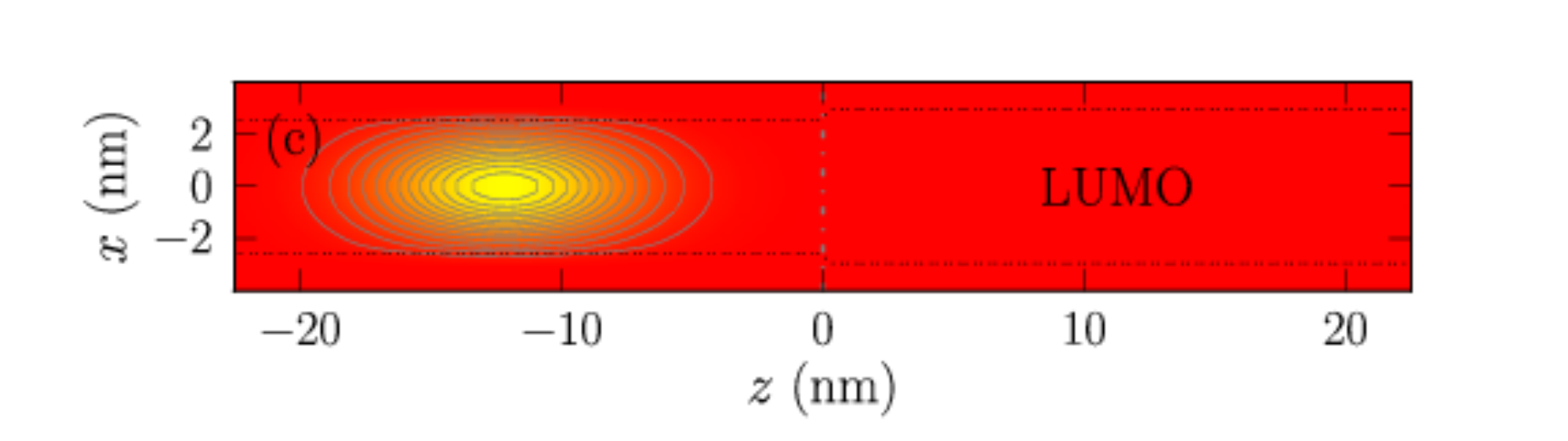}
\caption{\label{fig_wire_halfcs} a) VB edge along the axis of a GaAs nanowire which alternates segments with an AlAs shell and segments without a shell (core diameter $d_c=5$ nm, shell diameter $d_s=6$ nm; length of each segment $l=22.5$ nm). The potential would look like a smoothed $\simeq 0.35$ eV step function around each interface when $l\to\infty$ (same surface termination as in Fig.~\ref{fig_wire}). b) and c) 2D plots of the envelopes of the highest occupied and of the lowest unoccupied states, respectively.}
\end{figure}

In Fig.~\ref{fig_wire_halfcs}, we consider a GaAs nanowire which alternates segments with an AlAs shell and segments without a shell. This kind of heterostructure might be fabricated by AlAs overgrowth over a GaAs nanowire followed by a selective etching of the shell. This {\em inhomogeneous} shell does not fulfill the conditions of paragraph \ref{subsectionDisccoreshell}: the GaAs/AlAs interface is not a closed shape. We can therefore expect significant band edge modulations in the core. The potential in the core actually looks approximately like a sine function, because the surface/interface dipoles drive the band edges to different limits in each segment.  Its overall shape can also be inferred from Eq. (\ref{eqV1}). It would tend to a smoothed step function when $l\to\infty$, with a transition region around each shell extremity of width $\propto R$. Such structures efficiently separate carriers, the holes being localized in the segments with a shell, and the electrons in the segments without a shell. Therefore the separation of electron-hole pairs, which is highly desirable for, e.g., photovoltaics, does not necessarily require nanostructures with type-II interfaces but can be also obtained with type-I interfaces (Fig.~\ref{fig_wire_inv}), or even with homogeneous cores and appropriate passivation (Fig.~\ref{fig_wire_halfcs}), which would limit carrier diffusion. Large variations of the band edges can, in particular, be expected in core/shell nanowires with no common atom.\cite{Bauer83,Algra11} The sensitivity of nanowires to local modifications of their surface (even by neutral polar species) is also attractive for sensor applications.\cite{Xuan2010} Note that inhomogeneous shells or cappings are ubiquitous in nanowire devices, which can feature different materials or stacks of materials around in different parts of the device.

\subsection{Nanostructures with a mixed (0D/1D) dimensionality}
\label{subsectionMixed}

\subsubsection{Dot-rod heterostructures}

\begin{figure}
\includegraphics[width = .90\columnwidth]{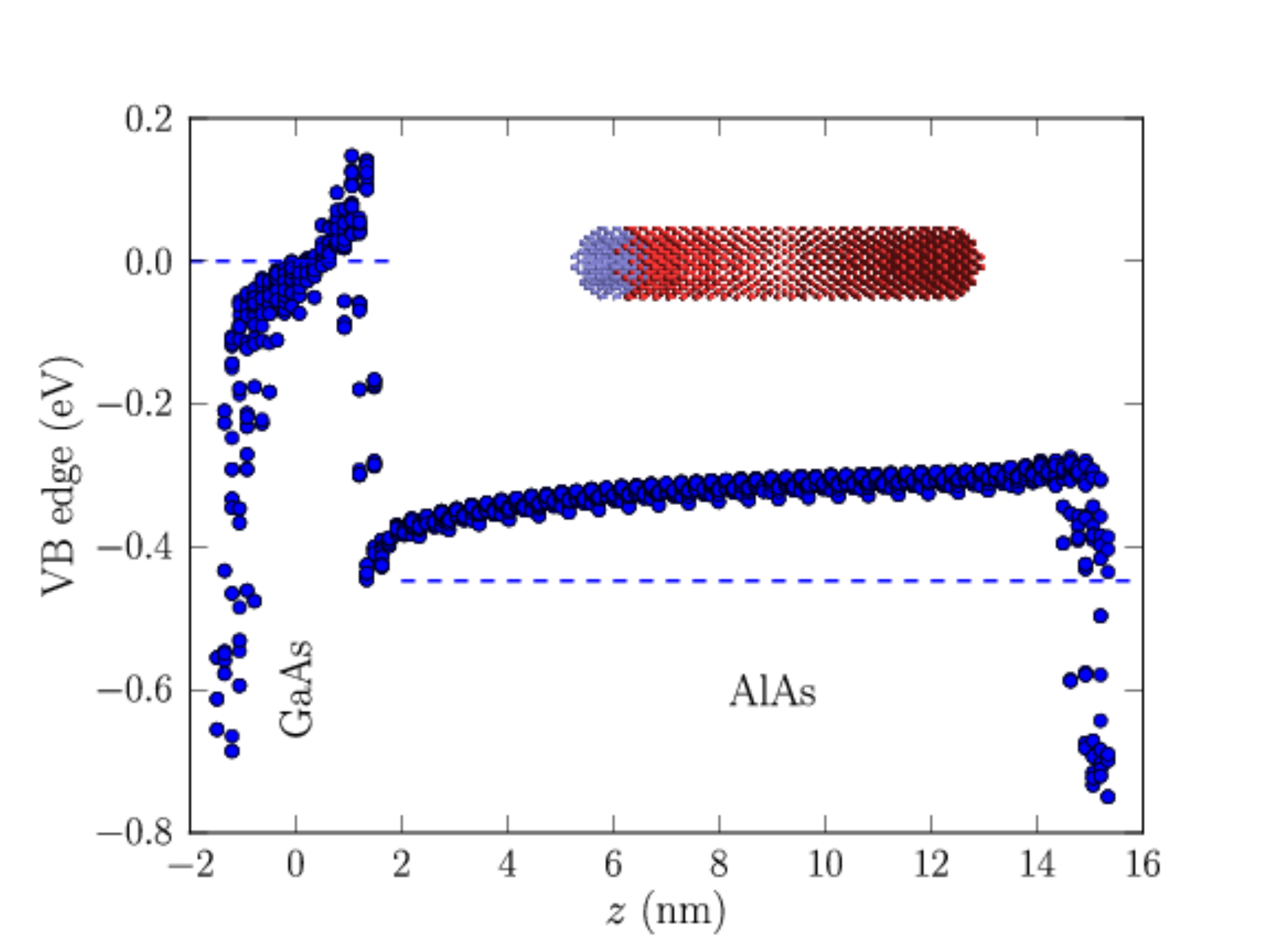}
\caption{\label{fig_dotrod} VB edge in a GaAs-quantum-dot/AlAs-quantum-rod heterostructure (diameter of the core sphere and of the cylindrical rod $d=3.2$ nm; total length $l=17.2$ nm). For clarity, only atoms with $r<1.2$ nm are plotted, where $r$ is the distance to the symmetry axis. The dashed horizontal lines indicate the VB offset in the 2D GaAs/AlAs super-lattice.}
\end{figure}

Figure~\ref{fig_dotrod} presents results for a nanostructure with a mixed (0D/1D) dimensionality. An AlAs quantum rod is attached to a spherical GaAs quantum dot. This type of structures presently receives considerable attention. \cite{Talapin03,Sadtler09,Borys10} As expected from nanowire super-lattices, the VB edge varies along the rod axis because it tends to the value for the pristine AlAs rod. Interestingly, the VB edge shows similar variation in the GaAs quantum dot in spite of its small size. Recently, Borys {\it et al.} \cite{Borys10} have shown that the interfacial energy transfer in CdSe/CdS heterostructure nanocrystals strongly depends on the particle morphology, for instance on the presence of CdS bulbs around the CdSe core in CdSe/CdS nanorods. Our results indeed suggest that the presence of these CdS bulbs could considerably modify the VB and CB edges in the nanostructures by changing the surface potential. As discussed above, the interfacial energy transfer could be also influenced by the presence of a barrier at the interface.

\subsubsection{Dumbbell quantum-dot heterostructures}

\begin{figure}
\includegraphics[width = .90\columnwidth]{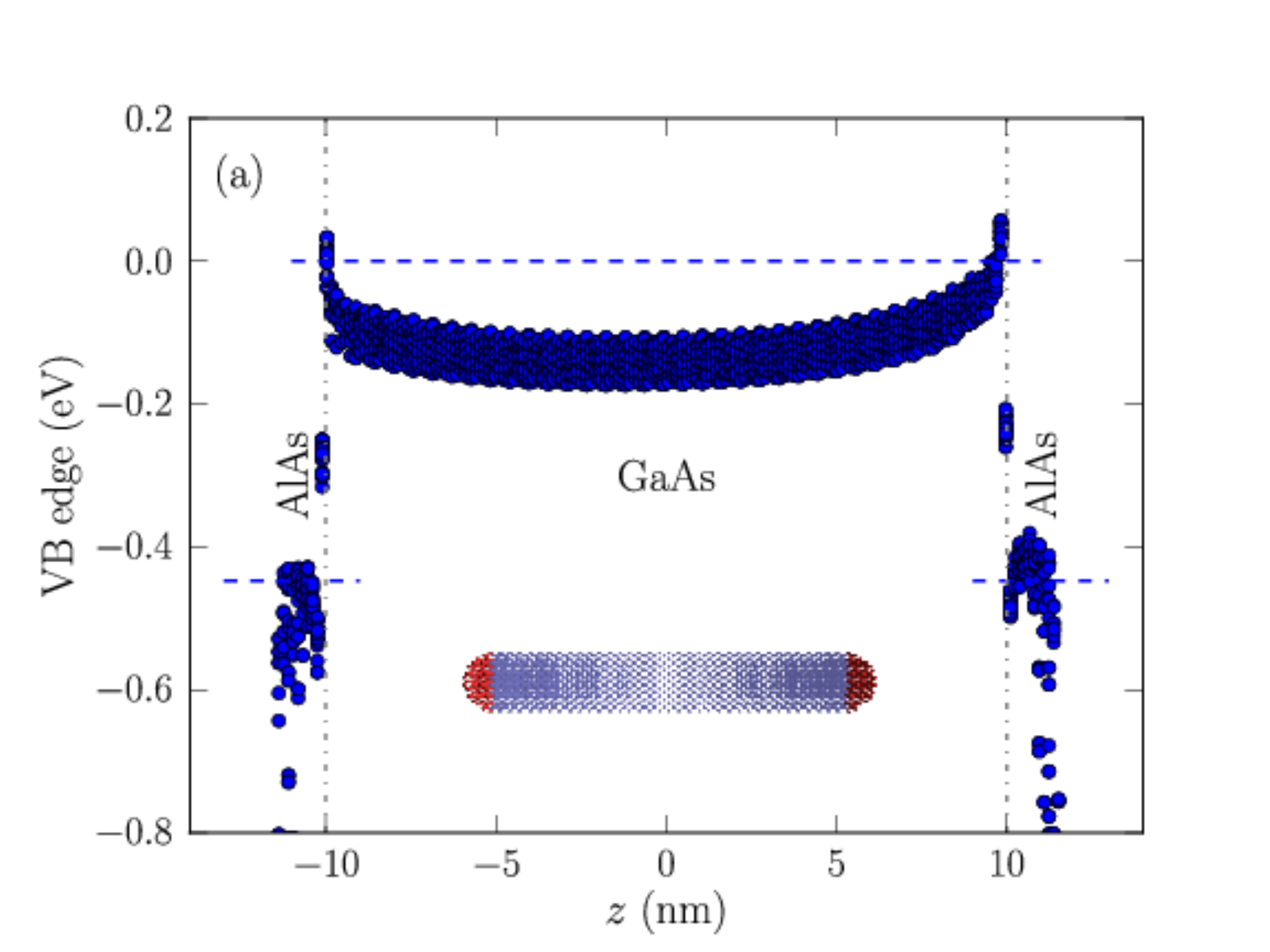}
\includegraphics[width = .90\columnwidth]{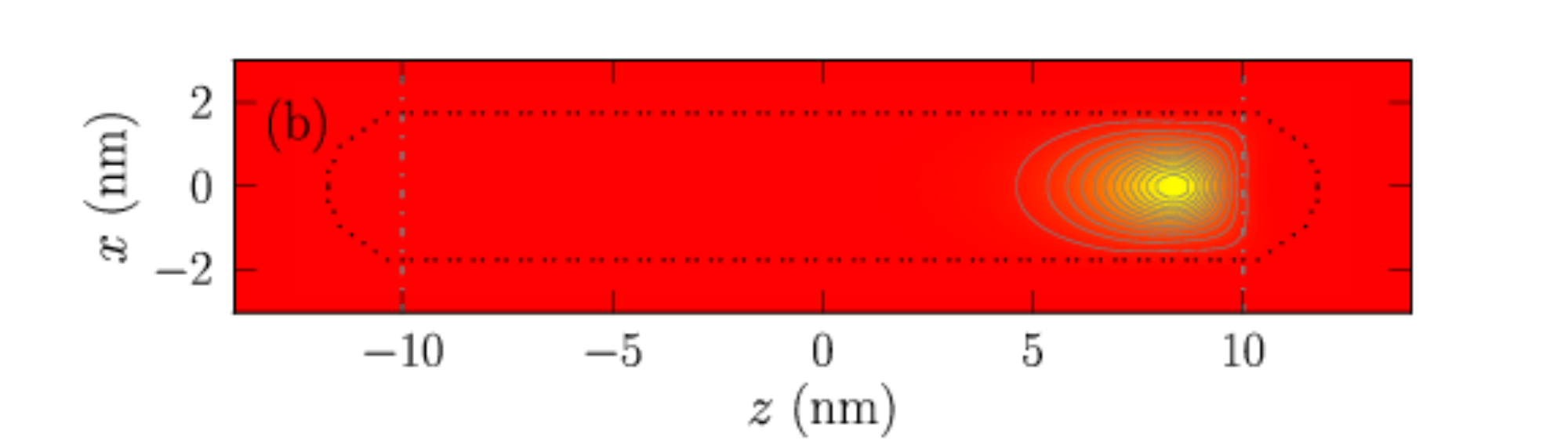}
\includegraphics[width = .90\columnwidth]{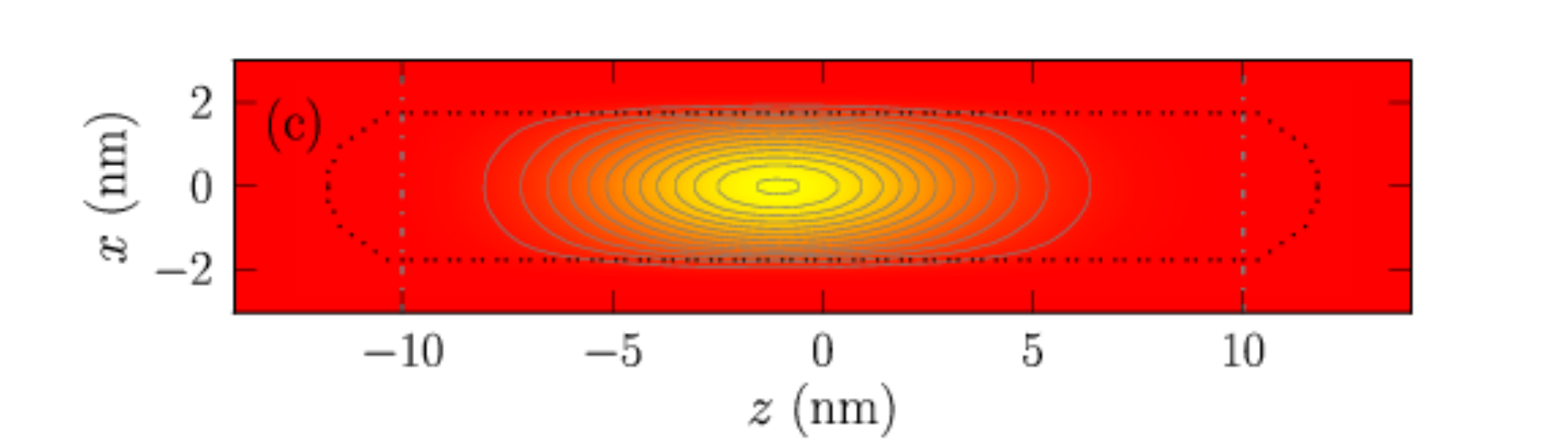}
\caption{\label{fig_dumbbell} a) VB edge along the axis of a GaAs/AlAs dumbbell quantum dot (diameter of the GaAs cylinder and of the AlAs hemispheres $d=3.4$ nm, length of the GaAs segment $l=20$ nm). b) and c) 2D plots of the envelopes of the highest occupied and of the lowest unoccupied states, respectively.}
\end{figure}

The synthesis of nanoscale heterostructures in the form of dumbbells has been recently reported.\cite{Kudera05} Figure~\ref{fig_dumbbell}a presents the VB edge in a cylindrical GaAs nanorod on which AlAs hemispheres have been grown at each tip. Altough they have a very small volume, the AlAs hemispheres have a profound influence on the potential in the GaAs rod. As a consequence, the highest occupied state is localized near the right interface (Fig.~\ref{fig_dumbbell}b) while the lowest unoccupied state is centered in the GaAs rod (Fig.~\ref{fig_dumbbell}c). The localization of the highest occupied level on the right side of the structure is due to a small asymmetry in the dumbbell. There is another state with almost the same energy on the left side.

\section{Conclusion}

In summary, we have presented self-consistent tight-binding calculations of nanoscale semiconductor heterojunctions. We have developed algorithms and methods allowing the study 0D and 1D nanostructures containing more than 75\,000 atoms per unit cell. Our analysis unveils the important physics which determines the band edges at the nanoscale. The band discontinuity right at the interfaces is found to be almost independent of the shape, dimensionality, and  size of the nano-objects. Hence the notion of band offset is robust in nanoscale heterojunctions, but the usual picture of square wells and barriers is only valid in particular cases. Indeed, in core-shell nanostructures with homogeneous shells, the band edges have a simple step-like behavior like in the 2D case. However, in nanowires or nanorods with axial heterostructures, the band edges have a more complex behavior which is strongly influenced by the nature of the surfaces (even if passivated). Indeed, the surface dipoles control the inner potential far from the interfaces, so that inhomogeneous surface termination leads to inhomogeneous band edge profiles which can trap or separate carriers. The electronic properties of such axial heterostructures can therefore be tailored by surface manipulations, which opens up new opportunities for the design, of e.g., nanoscale photovoltaics or sensor devices.

\begin{acknowledgments}
This work was supported by the French National Research Agency (ANR) projects QuantaMonde (contract ANR-07-NANO-023-02) and Quasanova (contract ANR-10-NANO-011-02), the EU Seventh Framework Program (EU-FP7 ITN Herodot), and by the French Ministry of Defense under grant no. 2008.34.0031. Part of the calculations were run at the CCRT supercomputing center. We thank H. Mariette, J. M. G\'erard, B. Grandidier and P. Caroff for a careful reading of the manuscript.
\end{acknowledgments}

\appendix*

\section{Computational details}

As usual in tight-binding, \cite{Delerue97,Delerue03} only the diagonal matrix elements of $V^{\rm sc}$ are considered. The potential on atom $i$ is written $V^{\rm sc}_{i}=-e\sum_{j}q_{j}/R_{ij}-V^{\rm mad}$, where $q_j$ is the charge on atom $j$ and $R_{ij}$ is the distance between atoms $i$ and $j$. We remove the bulk Madelung potential $V^{\rm mad}$ from $V^{\rm sc}_{i}$, because it is implicitely included in the tight-binding parameters ($H_0$) which account for the potential due to the ionic charges in bulk. In periodic systems, $V^{\rm sc}_{i}$ is calculated using the Ewald summation method.

The most common prescription for self-consistent calculations is to diagonalize the hamiltonian and compute the charge from the occupied states. This approach is however impractical, if not impossible for thousand atoms systems. We therefore compute the local density of states $\rho_{i}(E)=-{\rm Im} G_{ii}(E)/\pi$ from the green function $G(E)$ of the system, then use contour deformation techniques to integrate $\rho_{i}(z)$ in the complex plane. The net charge converges (up to $10^{-6}$ electrons) with only 48 points on the contour proposed in Ref.~\onlinecite{Brandbyge02}. The green function $G(z)$ is efficiently computed with the knitting algorithm of Ref.~\onlinecite{Kazymyrenko08}. The procedure can easily be parallelized and scales as $R^7$ (instead of $R^9$ for full diagonalization) in a nanocrystal with radius $R$, so that self-consistency can be achieved in just a few hours on 48 processors for tens of thousands of atoms.

The pseudo-hydrogen atoms are described by a single $s$ orbital with nearest-neighbor tight-binding parameters $V_{ss\sigma}=-4.00$ eV and $V_{sp\sigma}=6.93$ eV. The on-site $s$ orbital energy is equal to $-0.85$ eV, $0.65$ eV and $4.85$ eV for pseudo-hydrogen atoms nearest neighbor of Ga, Al and As atoms, respectively. The on-site parameters have been adjusted to minimize the charge transfers between pseudo-hydrogen and semiconductor atoms ($<0.1e$), while achieving a ionization potential $\simeq0.1$ eV higher in (large) AlAs nanocrystals than in GaAs nanocrystals. Some results presented in this paper have been obtained using different on-site $s$ orbital energies equal to $0.22$ eV and $-0.42$ eV for pseudo-hydrogen atoms nearest neighbor of Ga and Al atoms, respectively, which increases the ionization potential difference up to 0.8 eV. We emphasize that different sets of hydrogen parameters can reach the same difference of ionization potentials (with different charge transfers), but yield almost equivalent results. If the charge transfer between pseudo-hydrogen and semiconductor atoms becomes much larger, there is, as expected, stronger {\it lateral} band bending near the surface of the nanowires for example, but with similar variations of the potential {\it along} the nanowires.

\bibliography{offset_prb}

\providecommand{\noopsort}[1]{}\providecommand{\singleletter}[1]{#1}%
\begin{thebibliography}{10}%
\makeatletter
\providecommand \@ifxundefined [1]{%
 \ifx #1\undefined \expandafter \@firstoftwo
 \else \expandafter \@secondoftwo
\fi
}%
\providecommand \@ifnum [1]{%
 \ifnum #1\expandafter \@firstoftwo
 \else \expandafter \@secondoftwo
\fi
}%
\providecommand \enquote [1]{``#1''}%
\providecommand \bibnamefont  [1]{#1}%
\providecommand \bibfnamefont [1]{#1}%
\providecommand \citenamefont [1]{#1}%
\providecommand\href[0]{\@sanitize\@href}%
\providecommand\@href[1]{\endgroup\@@startlink{#1}\endgroup\@@href}%
\providecommand\@@href[1]{#1\@@endlink}%
\providecommand \@sanitize [0]{\begingroup\catcode`\&12\catcode`\#12\relax}%
\@ifxundefined \pdfoutput {\@firstoftwo}{%
 \@ifnum{\z@=\pdfoutput}{\@firstoftwo}{\@secondoftwo}%
}{%
 \providecommand\@@startlink[1]{\leavevmode\special{html:<a href="#1">}}%
 \providecommand\@@endlink[0]{\special{html:</a>}}%
}{%
 \providecommand\@@startlink[1]{%
  \leavevmode
  \pdfstartlink
   attr{/Border[0 0 1 ]/H/I/C[0 1 1]}%
   user{/Subtype/Link/A<</Type/Action/S/URI/URI(#1)>>}%
  \relax
 }%
 \providecommand\@@endlink[0]{\pdfendlink}%
}%
\providecommand \url  [0]{\begingroup\@sanitize \@url }%
\providecommand \@url [1]{\endgroup\@href {#1}{\urlprefix}}%
\providecommand \urlprefix [0]{URL }%
\providecommand \Eprint[0]{\href }%
\@ifxundefined \urlstyle {%
  \providecommand \doi [1]{doi:\discretionary{}{}{}#1}%
}{%
  \providecommand \doi [0]{doi:\discretionary{}{}{}\begingroup
  \urlstyle{rm}\Url }%
}%
\providecommand \doibase [0]{http://dx.doi.org/}%
\providecommand \Doi[1]{\href{\doibase#1}}%
\providecommand \bibAnnote [3]{%
  \BibitemShut{#1}%
  \begin{quotation}\noindent
    \textsc{Key:}\ #2\\\textsc{Annotation:}\ #3%
  \end{quotation}%
}%
\providecommand \bibAnnoteFile [2]{%
  \IfFileExists{#2}{\bibAnnote {#1} {#2} {\input{#2}}}{}%
}%
\providecommand \typeout [0]{\immediate \write \m@ne }%
\providecommand \selectlanguage [0]{\@gobble}%
\providecommand \bibinfo [0]{\@secondoftwo}%
\providecommand \bibfield [0]{\@secondoftwo}%
\providecommand \translation [1]{[#1]}%
\providecommand \BibitemOpen[0]{}%
\providecommand \bibitemStop [0]{}%
\providecommand \bibitemNoStop [0]{.\EOS\space}%
\providecommand \EOS [0]{\spacefactor3000\relax}%
\providecommand \BibitemShut [1]{\csname bibitem#1\endcsname}%
\bibitem{Bastard88}%
  \BibitemOpen
  \bibfield{author}{%
  \bibinfo {author} {\bibfnamefont{G.}~\bibnamefont{Bastard}},\ }%
  \emph{\bibinfo {title} {Wave Mechanics Applied to Semiconductor
  Heterostructures}}\ (\bibinfo {publisher} {Editions de Physique},\ \bibinfo
  {address} {Les Ulis, France},\ \bibinfo {year} {1988})%
  \bibAnnoteFile{NoStop}{Bastard88}%
\bibitem{Burt92}%
  \BibitemOpen
  \bibfield{author}{%
  \bibinfo {author} {\bibfnamefont{M.~G.}\ \bibnamefont{Burt}},\ }%
  \bibfield{journal}{%
  \bibinfo {journal} {J. Phys. Condens. Matter}\ }%
  \textbf{\bibinfo {volume} {4}},\ \bibinfo {pages} {6651} (\bibinfo {year}
  {1992}),\ \url{http://stacks.iop.org/0953-8984/4/i=32/a=003}%
  \bibAnnoteFile{NoStop}{Burt92}%
\bibitem{Vurgaftman01}%
  \BibitemOpen
  \bibfield{author}{%
  \bibinfo {author} {\bibfnamefont{I.}~\bibnamefont{Vurgaftman}}, \bibinfo
  {author} {\bibfnamefont{J.~R.}\ \bibnamefont{Meyer}},\ and\ \bibinfo {author}
  {\bibfnamefont{L.~R.}\ \bibnamefont{Ram-Mohan}},\ }%
  \bibfield{journal}{%
  \bibinfo {journal} {J. Appl. Phys.}\ }%
  \textbf{\bibinfo {volume} {89}},\ \bibinfo {pages} {5815} (\bibinfo {year}
  {2001})%
  \bibAnnoteFile{NoStop}{Vurgaftman01}%
\bibitem{Peng97}%
  \BibitemOpen
  \bibfield{author}{%
  \bibinfo {author} {\bibfnamefont{X.}~\bibnamefont{Peng}}, \bibinfo {author}
  {\bibfnamefont{M.~C.}\ \bibnamefont{Schlamp}}, \bibinfo {author}
  {\bibfnamefont{A.~V.}\ \bibnamefont{Kadavanich}},\ and\ \bibinfo {author}
  {\bibfnamefont{A.~P.}\ \bibnamefont{Alivisatos}},\ }%
  \bibfield{journal}{%
  \Doi{10.1021/ja970754m}{\bibinfo {journal} {J. Am. Chem. Soc.}}\ }%
  \textbf{\bibinfo {volume} {119}},\ \bibinfo {pages} {7019} (\bibinfo {year}
  {1997}),\ \url{http://pubs.acs.org/doi/abs/10.1021/ja970754m}%
  \bibAnnoteFile{NoStop}{Peng97}%
\bibitem{Dabboussi97}%
  \BibitemOpen
  \bibfield{author}{%
  \bibinfo {author} {\bibfnamefont{B.~O.}\ \bibnamefont{Dabbousi}}, \bibinfo
  {author} {\bibfnamefont{J.}~\bibnamefont{Rodriguez-Viejo}}, \bibinfo {author}
  {\bibfnamefont{F.~V.}\ \bibnamefont{Mikulec}}, \bibinfo {author}
  {\bibfnamefont{J.~R.}\ \bibnamefont{Heine}}, \bibinfo {author}
  {\bibfnamefont{H.}~\bibnamefont{Mattoussi}}, \bibinfo {author}
  {\bibfnamefont{R.}~\bibnamefont{Ober}}, \bibinfo {author}
  {\bibfnamefont{K.~F.}\ \bibnamefont{Jensen}},\ and\ \bibinfo {author}
  {\bibfnamefont{M.~G.}\ \bibnamefont{Bawendi}},\ }%
  \bibfield{journal}{%
  \Doi{10.1021/jp971091y}{\bibinfo {journal} {J. Phys. Chem. B}}\ }%
  \textbf{\bibinfo {volume} {101}},\ \bibinfo {pages} {9463} (\bibinfo {year}
  {1997}),\ \url{http://pubs.acs.org/doi/abs/10.1021/jp971091y}%
  \bibAnnoteFile{NoStop}{Dabboussi97}%
\bibitem{Mekis03}%
  \BibitemOpen
  \bibfield{author}{%
  \bibinfo {author} {\bibfnamefont{I.}~\bibnamefont{Mekis}}, \bibinfo {author}
  {\bibfnamefont{D.~V.}\ \bibnamefont{Talapin}}, \bibinfo {author}
  {\bibfnamefont{A.}~\bibnamefont{Kornowski}}, \bibinfo {author}
  {\bibfnamefont{M.}~\bibnamefont{Haase}},\ and\ \bibinfo {author}
  {\bibfnamefont{H.}~\bibnamefont{Weller}},\ }%
  \bibfield{journal}{%
  \Doi{10.1021/jp0278364}{\bibinfo {journal} {J. Phys. Chem. B}}\ }%
  \textbf{\bibinfo {volume} {107}},\ \bibinfo {pages} {7454} (\bibinfo {year}
  {2003}),\ \url{http://pubs.acs.org/doi/abs/10.1021/jp0278364}%
  \bibAnnoteFile{NoStop}{Mekis03}%
\bibitem{Mahler08}%
  \BibitemOpen
  \bibfield{author}{%
  \bibinfo {author} {\bibfnamefont{B.}~\bibnamefont{Mahler}}, \bibinfo {author}
  {\bibfnamefont{P.}~\bibnamefont{Spinicelli}}, \bibinfo {author}
  {\bibfnamefont{S.}~\bibnamefont{Buil}}, \bibinfo {author}
  {\bibfnamefont{X.}~\bibnamefont{Quelin}}, \bibinfo {author}
  {\bibfnamefont{J.-P.}\ \bibnamefont{Hermier}},\ and\ \bibinfo {author}
  {\bibfnamefont{B.}~\bibnamefont{Dubertret}},\ }%
  \bibfield{journal}{%
  \Doi{10.1038/nmat2222}{\bibinfo {journal} {Nature Mater.}}\ }%
  \textbf{\bibinfo {volume} {7}},\ \bibinfo {pages} {659} (\bibinfo {year}
  {2008}),\ \url{http://dx.doi.org/10.1038/nmat2222}%
  \bibAnnoteFile{NoStop}{Mahler08}%
\bibitem{Talapin03}%
  \BibitemOpen
  \bibfield{author}{%
  \bibinfo {author} {\bibfnamefont{D.~V.}\ \bibnamefont{Talapin}}, \bibinfo
  {author} {\bibfnamefont{R.}~\bibnamefont{Koeppe}}, \bibinfo {author}
  {\bibfnamefont{S.}~\bibnamefont{Götzinger}}, \bibinfo {author}
  {\bibfnamefont{A.}~\bibnamefont{Kornowski}}, \bibinfo {author}
  {\bibfnamefont{J.~M.}\ \bibnamefont{Lupton}}, \bibinfo {author}
  {\bibfnamefont{A.~L.}\ \bibnamefont{Rogach}}, \bibinfo {author}
  {\bibfnamefont{O.}~\bibnamefont{Benson}}, \bibinfo {author}
  {\bibfnamefont{J.}~\bibnamefont{Feldmann}},\ and\ \bibinfo {author}
  {\bibfnamefont{H.}~\bibnamefont{Weller}},\ }%
  \bibfield{journal}{%
  \Doi{10.1021/nl034815s}{\bibinfo {journal} {Nano Lett.}}\ }%
  \textbf{\bibinfo {volume} {3}},\ \bibinfo {pages} {1677} (\bibinfo {year}
  {2003}),\ \url{http://pubs.acs.org/doi/abs/10.1021/nl034815s}%
  \bibAnnoteFile{NoStop}{Talapin03}%
\bibitem{Sadtler09}%
  \BibitemOpen
  \bibfield{author}{%
  \bibinfo {author} {\bibfnamefont{B.}~\bibnamefont{Sadtler}}, \bibinfo
  {author} {\bibfnamefont{D.~O.}\ \bibnamefont{Demchenko}}, \bibinfo {author}
  {\bibfnamefont{H.}~\bibnamefont{Zheng}}, \bibinfo {author}
  {\bibfnamefont{S.~M.}\ \bibnamefont{Hughes}}, \bibinfo {author}
  {\bibfnamefont{M.~G.}\ \bibnamefont{Merkle}}, \bibinfo {author}
  {\bibfnamefont{U.}~\bibnamefont{Dahmen}}, \bibinfo {author}
  {\bibfnamefont{L.-W.}\ \bibnamefont{Wang}},\ and\ \bibinfo {author}
  {\bibfnamefont{A.~P.}\ \bibnamefont{Alivisatos}},\ }%
  \bibfield{journal}{%
  \Doi{10.1021/ja809854q}{\bibinfo {journal} {J. Am. Chem. Soc.}}\ }%
  \textbf{\bibinfo {volume} {131}},\ \bibinfo {pages} {5285} (\bibinfo {year}
  {2009}),\ \url{http://pubs.acs.org/doi/abs/10.1021/ja809854q}%
  \bibAnnoteFile{NoStop}{Sadtler09}%
\bibitem{Borys10}%
  \BibitemOpen
  \bibfield{author}{%
  \bibinfo {author} {\bibfnamefont{N.~J.}\ \bibnamefont{Borys}}, \bibinfo
  {author} {\bibfnamefont{M.~J.}\ \bibnamefont{Walter}}, \bibinfo {author}
  {\bibfnamefont{J.}~\bibnamefont{Huang}}, \bibinfo {author}
  {\bibfnamefont{D.~V.}\ \bibnamefont{Talapin}},\ and\ \bibinfo {author}
  {\bibfnamefont{J.~M.}\ \bibnamefont{Lupton}},\ }%
  \bibfield{journal}{%
  \Doi{10.1126/science.1198070}{\bibinfo {journal} {Science}}\ }%
  \textbf{\bibinfo {volume} {330}},\ \bibinfo {pages} {1371} (\bibinfo {year}
  {2010}),\ \url{http://www.sciencemag.org/content/330/6009/1371.abstract}%
  \bibAnnoteFile{NoStop}{Borys10}%
\bibitem{Chin07}%
  \BibitemOpen
  \bibfield{author}{%
  \bibinfo {author} {\bibfnamefont{P.~T.~K.}\ \bibnamefont{Chin}}, \bibinfo
  {author} {\bibfnamefont{C.}~\bibnamefont{de~Mello~Donegá}}, \bibinfo
  {author} {\bibfnamefont{S.~S.}\ \bibnamefont{van Bavel}}, \bibinfo {author}
  {\bibfnamefont{S.~C.~J.}\ \bibnamefont{Meskers}}, \bibinfo {author}
  {\bibfnamefont{N.~A. J.~M.}\ \bibnamefont{Sommerdijk}},\ and\ \bibinfo
  {author} {\bibfnamefont{R.~A.~J.}\ \bibnamefont{Janssen}},\ }%
  \bibfield{journal}{%
  \Doi{10.1021/ja0738071}{\bibinfo {journal} {J. Am. Chem. Soc.}}\ }%
  \textbf{\bibinfo {volume} {129}},\ \bibinfo {pages} {14880} (\bibinfo {year}
  {2007}),\ \url{http://pubs.acs.org/doi/abs/10.1021/ja0738071}%
  \bibAnnoteFile{NoStop}{Chin07}%
\bibitem{Talapin07}%
  \BibitemOpen
  \bibfield{author}{%
  \bibinfo {author} {\bibfnamefont{D.~V.}\ \bibnamefont{Talapin}}, \bibinfo
  {author} {\bibfnamefont{J.~H.}\ \bibnamefont{Nelson}}, \bibinfo {author}
  {\bibfnamefont{E.~V.}\ \bibnamefont{Shevchenko}}, \bibinfo {author}
  {\bibfnamefont{S.}~\bibnamefont{Aloni}}, \bibinfo {author}
  {\bibfnamefont{B.}~\bibnamefont{Sadtler}},\ and\ \bibinfo {author}
  {\bibfnamefont{A.~P.}\ \bibnamefont{Alivisatos}},\ }%
  \bibfield{journal}{%
  \Doi{10.1021/nl072003g}{\bibinfo {journal} {Nano Lett.}}\ }%
  \textbf{\bibinfo {volume} {7}},\ \bibinfo {pages} {2951} (\bibinfo {year}
  {2007}),\ \url{http://pubs.acs.org/doi/abs/10.1021/nl072003g}%
  \bibAnnoteFile{NoStop}{Talapin07}%
\bibitem{Kudera05}%
  \BibitemOpen
  \bibfield{author}{%
  \bibinfo {author} {\bibfnamefont{S.}~\bibnamefont{Kudera}}, \bibinfo {author}
  {\bibfnamefont{L.}~\bibnamefont{Carbone}}, \bibinfo {author}
  {\bibfnamefont{M.~F.}\ \bibnamefont{Casula}}, \bibinfo {author}
  {\bibfnamefont{R.}~\bibnamefont{Cingolani}}, \bibinfo {author}
  {\bibfnamefont{A.}~\bibnamefont{Falqui}}, \bibinfo {author}
  {\bibfnamefont{E.}~\bibnamefont{Snoeck}}, \bibinfo {author}
  {\bibfnamefont{W.~J.}\ \bibnamefont{Parak}},\ and\ \bibinfo {author}
  {\bibfnamefont{L.}~\bibnamefont{Manna}},\ }%
  \bibfield{journal}{%
  \Doi{10.1021/nl048060g}{\bibinfo {journal} {Nano Lett.}}\ }%
  \textbf{\bibinfo {volume} {5}},\ \bibinfo {pages} {445} (\bibinfo {year}
  {2005}),\ \url{http://pubs.acs.org/doi/abs/10.1021/nl048060g}%
  \bibAnnoteFile{NoStop}{Kudera05}%
\bibitem{Lee02}%
  \BibitemOpen
  \bibfield{author}{%
  \bibinfo {author} {\bibfnamefont{S.-M.}\ \bibnamefont{Lee}}, \bibinfo
  {author} {\bibfnamefont{Y.-w.}\ \bibnamefont{Jun}}, \bibinfo {author}
  {\bibfnamefont{S.-N.}\ \bibnamefont{Cho}},\ and\ \bibinfo {author}
  {\bibfnamefont{J.}~\bibnamefont{Cheon}},\ }%
  \bibfield{journal}{%
  \Doi{10.1021/ja026805j}{\bibinfo {journal} {J. Am. Chem. Soc.}}\ }%
  \textbf{\bibinfo {volume} {124}},\ \bibinfo {pages} {11244} (\bibinfo {year}
  {2002}),\ \url{http://pubs.acs.org/doi/abs/10.1021/ja026805j}%
  \bibAnnoteFile{NoStop}{Lee02}%
\bibitem{Bjork02}%
  \BibitemOpen
  \bibfield{author}{%
  \bibinfo {author} {\bibfnamefont{M.~T.}\ \bibnamefont{Bj\"ork}}, \bibinfo
  {author} {\bibfnamefont{B.~J.}\ \bibnamefont{Ohlsson}}, \bibinfo {author}
  {\bibfnamefont{T.}~\bibnamefont{Sass}}, \bibinfo {author}
  {\bibfnamefont{A.~I.}\ \bibnamefont{Persson}}, \bibinfo {author}
  {\bibfnamefont{C.}~\bibnamefont{Thelander}}, \bibinfo {author}
  {\bibfnamefont{M.~H.}\ \bibnamefont{Magnusson}}, \bibinfo {author}
  {\bibfnamefont{K.}~\bibnamefont{Deppert}}, \bibinfo {author}
  {\bibfnamefont{L.~R.}\ \bibnamefont{Wallenberg}},\ and\ \bibinfo {author}
  {\bibfnamefont{L.}~\bibnamefont{Samuelson}},\ }%
  \bibfield{journal}{%
  \Doi{10.1063/1.1447312}{\bibinfo {journal} {Appl. Phys. Lett.}}\ }%
  \textbf{\bibinfo {volume} {80}},\ \bibinfo {pages} {1058} (\bibinfo {year}
  {2002}),\ \url{http://link.aip.org/link/?APL/80/1058/1}%
  \bibAnnoteFile{NoStop}{Bjork02}%
\bibitem{Gudiksen02}%
  \BibitemOpen
  \bibfield{author}{%
  \bibinfo {author} {\bibfnamefont{M.~S.}\ \bibnamefont{Gudiksen}}, \bibinfo
  {author} {\bibfnamefont{L.~J.}\ \bibnamefont{Lauhon}}, \bibinfo {author}
  {\bibfnamefont{J.}~\bibnamefont{Wang}}, \bibinfo {author}
  {\bibfnamefont{D.~C.}\ \bibnamefont{Smith}},\ and\ \bibinfo {author}
  {\bibfnamefont{C.~M.}\ \bibnamefont{Lieber}},\ }%
  \bibfield{journal}{%
  \Doi{10.1038/415617a}{\bibinfo {journal} {Nature}}\ }%
  \textbf{\bibinfo {volume} {415}},\ \bibinfo {pages} {617} (\bibinfo {year}
  {2002})%
  \bibAnnoteFile{NoStop}{Gudiksen02}%
\bibitem{Dick10}%
  \BibitemOpen
  \bibfield{author}{%
  \bibinfo {author} {\bibfnamefont{K.~A.}\ \bibnamefont{Dick}}, \bibinfo
  {author} {\bibfnamefont{C.}~\bibnamefont{Thelander}}, \bibinfo {author}
  {\bibfnamefont{L.}~\bibnamefont{Samuelson}},\ and\ \bibinfo {author}
  {\bibfnamefont{P.}~\bibnamefont{Caroff}},\ }%
  \bibfield{journal}{%
  \Doi{10.1021/nl101632a}{\bibinfo {journal} {Nano Lett.}}\ }%
  \textbf{\bibinfo {volume} {10}},\ \bibinfo {pages} {3494} (\bibinfo {year}
  {2010})%
  \bibAnnoteFile{NoStop}{Dick10}%
\bibitem{Carnevale11}%
  \BibitemOpen
  \bibfield{author}{%
  \bibinfo {author} {\bibfnamefont{S.~D.}\ \bibnamefont{Carnevale}}, \bibinfo
  {author} {\bibfnamefont{J.}~\bibnamefont{Yang}}, \bibinfo {author}
  {\bibfnamefont{P.~J.}\ \bibnamefont{Phillips}}, \bibinfo {author}
  {\bibfnamefont{M.~J.}\ \bibnamefont{Mills}},\ and\ \bibinfo {author}
  {\bibfnamefont{R.~C.}\ \bibnamefont{Myers}},\ }%
  \bibfield{journal}{%
  \Doi{10.1021/nl104265u}{\bibinfo {journal} {Nano Lett.}}\ }%
  \textbf{\bibinfo {volume} {11}},\ \bibinfo {pages} {866} (\bibinfo {year}
  {2011})%
  \bibAnnoteFile{NoStop}{Carnevale11}%
\bibitem{Lauhon02}%
  \BibitemOpen
  \bibfield{author}{%
  \bibinfo {author} {\bibfnamefont{L.}~\bibnamefont{Lauhon}}, \bibinfo {author}
  {\bibfnamefont{M.}~\bibnamefont{Gudiksen}}, \bibinfo {author}
  {\bibfnamefont{C.}~\bibnamefont{Wang}},\ and\ \bibinfo {author}
  {\bibfnamefont{C.}~\bibnamefont{Lieber}},\ }%
  \bibfield{journal}{%
  \Doi{10.1038/nature01141}{\bibinfo {journal} {Nature}}\ }%
  \textbf{\bibinfo {volume} {420}},\ \bibinfo {pages} {57} (\bibinfo {year}
  {2002})%
  \bibAnnoteFile{NoStop}{Lauhon02}%
\bibitem{Noborisaka05}%
  \BibitemOpen
  \bibfield{author}{%
  \bibinfo {author} {\bibfnamefont{J.}~\bibnamefont{Noborisaka}}, \bibinfo
  {author} {\bibfnamefont{J.}~\bibnamefont{Motohisa}}, \bibinfo {author}
  {\bibfnamefont{S.}~\bibnamefont{Hara}},\ and\ \bibinfo {author}
  {\bibfnamefont{T.}~\bibnamefont{Fukui}},\ }%
  \bibfield{journal}{%
  \Doi{10.1063/1.2035332}{\bibinfo {journal} {Appl. Phys. Lett.}}\ }%
  \textbf{\bibinfo {volume} {87}},\ \bibinfo {pages} {093109} (\bibinfo {year}
  {2005}),\ \url{http://link.aip.org/link/?APL/87/093109/1}%
  \bibAnnoteFile{NoStop}{Noborisaka05}%
\bibitem{Skold05}%
  \BibitemOpen
  \bibfield{author}{%
  \bibinfo {author} {\bibfnamefont{N.}~\bibnamefont{Sk\"old}}, \bibinfo
  {author} {\bibfnamefont{L.~S.}\ \bibnamefont{Karlsson}}, \bibinfo {author}
  {\bibfnamefont{M.~W.}\ \bibnamefont{Larsson}}, \bibinfo {author}
  {\bibfnamefont{M.-E.}\ \bibnamefont{Pistol}}, \bibinfo {author}
  {\bibfnamefont{W.}~\bibnamefont{Seifert}}, \bibinfo {author}
  {\bibfnamefont{J.}~\bibnamefont{Tr\"agardh}},\ and\ \bibinfo {author}
  {\bibfnamefont{L.}~\bibnamefont{Samuelson}},\ }%
  \bibfield{journal}{%
  \Doi{10.1021/nl051304s}{\bibinfo {journal} {Nano Lett.}}\ }%
  \textbf{\bibinfo {volume} {5}},\ \bibinfo {pages} {1943} (\bibinfo {year}
  {2005}),\ \url{http://pubs.acs.org/doi/abs/10.1021/nl051304s}%
  \bibAnnoteFile{NoStop}{Skold05}%
\bibitem{Note_relaxations}%
  \BibitemOpen
  \bibinfo {note} {The effect of strains is usually superimposed to the
  potential.}%
  \bibAnnoteFile{Stop}{Note_relaxations}%
\bibitem{Haus93}%
  \BibitemOpen
  \bibfield{author}{%
  \bibinfo {author} {\bibfnamefont{J.~W.}\ \bibnamefont{Haus}}, \bibinfo
  {author} {\bibfnamefont{H.~S.}\ \bibnamefont{Zhou}}, \bibinfo {author}
  {\bibfnamefont{I.}~\bibnamefont{Honma}},\ and\ \bibinfo {author}
  {\bibfnamefont{H.}~\bibnamefont{Komiyama}},\ }%
  \bibfield{journal}{%
  \Doi{10.1103/PhysRevB.47.1359}{\bibinfo {journal} {Phys. Rev. B}}\ }%
  \textbf{\bibinfo {volume} {47}},\ \bibinfo {pages} {1359} (\bibinfo {month}
  {Jan}\ \bibinfo {year} {1993})%
  \bibAnnoteFile{NoStop}{Haus93}%
\bibitem{Garcia08}%
  \BibitemOpen
  \bibfield{author}{%
  \bibinfo {author} {\bibfnamefont{L.~F.}\ \bibnamefont{Garcia}}, \bibinfo
  {author} {\bibfnamefont{J.}~\bibnamefont{Silva-Valencia}}, \bibinfo {author}
  {\bibfnamefont{I.~D.}\ \bibnamefont{MikhailoVa}},\ and\ \bibinfo {author}
  {\bibfnamefont{J.~E.}\ \bibnamefont{Galvan-Moya}},\ }%
  \bibfield{journal}{%
  \Doi{10.1016/j.physb.2007.06.007}{\bibinfo {journal} {Physica B}}\ }%
  \textbf{\bibinfo {volume} {403}},\ \bibinfo {pages} {5} (\bibinfo {month}
  {JAN 1}\ \bibinfo {year} {2008})%
  \bibAnnoteFile{NoStop}{Garcia08}%
\bibitem{Pistol09}%
  \BibitemOpen
  \bibfield{author}{%
  \bibinfo {author} {\bibfnamefont{M.-E.}\ \bibnamefont{Pistol}}\ and\ \bibinfo
  {author} {\bibfnamefont{C.~E.}\ \bibnamefont{Pryor}},\ }%
  \bibfield{journal}{%
  \Doi{10.1103/PhysRevB.80.035316}{\bibinfo {journal} {Phys. Rev. B}}\ }%
  \textbf{\bibinfo {volume} {80}},\ \bibinfo {pages} {035316} (\bibinfo {month}
  {Jul}\ \bibinfo {year} {2009})%
  \bibAnnoteFile{NoStop}{Pistol09}%
\bibitem{Lue09}%
  \BibitemOpen
  \bibfield{author}{%
  \bibinfo {author} {\bibfnamefont{X.}~\bibnamefont{Lue}},\ }%
  \bibfield{journal}{%
  \Doi{10.1063/1.3223329}{\bibinfo {journal} {J. Appl. Phys.}}\ }%
  \textbf{\bibinfo {volume} {106}},\ \bibinfo {pages} {064305} (\bibinfo
  {month} {SEP 15}\ \bibinfo {year} {2009})%
  \bibAnnoteFile{NoStop}{Lue09}%
\bibitem{Perez03}%
  \BibitemOpen
  \bibfield{author}{%
  \bibinfo {author} {\bibfnamefont{J.}~\bibnamefont{P\'erez-Conde}}\ and\
  \bibinfo {author} {\bibfnamefont{A.~K.}\ \bibnamefont{Bhattacharjee}},\ }%
  \bibfield{journal}{%
  \Doi{10.1103/PhysRevB.67.235303}{\bibinfo {journal} {Phys. Rev. B}}\ }%
  \textbf{\bibinfo {volume} {67}},\ \bibinfo {pages} {235303} (\bibinfo {month}
  {Jun}\ \bibinfo {year} {2003})%
  \bibAnnoteFile{NoStop}{Perez03}%
\bibitem{Diaz06}%
  \BibitemOpen
  \bibfield{author}{%
  \bibinfo {author} {\bibfnamefont{J.~G.}\ \bibnamefont{D\'\i{}az}}, \bibinfo
  {author} {\bibfnamefont{M.}~\bibnamefont{Zieli\ifmmode~\acute{n}\else
  \'{n}\fi{}ski}}, \bibinfo {author}
  {\bibfnamefont{W.}~\bibnamefont{Jask\'olski}},\ and\ \bibinfo {author}
  {\bibfnamefont{G.~W.}\ \bibnamefont{Bryant}},\ }%
  \bibfield{journal}{%
  \Doi{10.1103/PhysRevB.74.205309}{\bibinfo {journal} {Phys. Rev. B}}\ }%
  \textbf{\bibinfo {volume} {74}},\ \bibinfo {pages} {205309} (\bibinfo {month}
  {Nov}\ \bibinfo {year} {2006})%
  \bibAnnoteFile{NoStop}{Diaz06}%
\bibitem{Niquet08}%
  \BibitemOpen
  \bibfield{author}{%
  \bibinfo {author} {\bibfnamefont{Y.-M.}\ \bibnamefont{Niquet}}\ and\ \bibinfo
  {author} {\bibfnamefont{D.~C.}\ \bibnamefont{Mojica}},\ }%
  \bibfield{journal}{%
  \Doi{10.1103/PhysRevB.77.115316}{\bibinfo {journal} {Phys. Rev. B}}\ }%
  \textbf{\bibinfo {volume} {77}},\ \bibinfo {pages} {115316} (\bibinfo {month}
  {Mar}\ \bibinfo {year} {2008})%
  \bibAnnoteFile{NoStop}{Niquet08}%
\bibitem{Schrier06}%
  \BibitemOpen
  \bibfield{author}{%
  \bibinfo {author} {\bibfnamefont{J.}~\bibnamefont{Schrier}}\ and\ \bibinfo
  {author} {\bibfnamefont{L.-W.}\ \bibnamefont{Wang}},\ }%
  \bibfield{journal}{%
  \Doi{10.1103/PhysRevB.73.245332}{\bibinfo {journal} {Phys. Rev. B}}\ }%
  \textbf{\bibinfo {volume} {73}},\ \bibinfo {pages} {245332} (\bibinfo {month}
  {Jun}\ \bibinfo {year} {2006})%
  \bibAnnoteFile{NoStop}{Schrier06}%
\bibitem{Zhang10}%
  \BibitemOpen
  \bibfield{author}{%
  \bibinfo {author} {\bibfnamefont{L.}~\bibnamefont{Zhang}}, \bibinfo {author}
  {\bibfnamefont{J.-W.}\ \bibnamefont{Luo}}, \bibinfo {author}
  {\bibfnamefont{A.}~\bibnamefont{Zunger}}, \bibinfo {author}
  {\bibfnamefont{N.}~\bibnamefont{Akopian}}, \bibinfo {author}
  {\bibfnamefont{V.}~\bibnamefont{Zwiller}},\ and\ \bibinfo {author}
  {\bibfnamefont{J.-C.}\ \bibnamefont{Harmand}},\ }%
  \bibfield{journal}{%
  \Doi{10.1021/nl102109s}{\bibinfo {journal} {Nano Lett.}}\ }%
  \textbf{\bibinfo {volume} {10}},\ \bibinfo {pages} {4055} (\bibinfo {month}
  {OCT}\ \bibinfo {year} {2010})%
  \bibAnnoteFile{NoStop}{Zhang10}%
\bibitem{Leonard00}%
  \BibitemOpen
  \bibfield{author}{%
  \bibinfo {author} {\bibfnamefont{F.}~\bibnamefont{L\'eonard}}\ and\ \bibinfo
  {author} {\bibfnamefont{J.}~\bibnamefont{Tersoff}},\ }%
  \bibfield{journal}{%
  \Doi{10.1103/PhysRevLett.84.4693}{\bibinfo {journal} {Phys. Rev. Lett.}}\ }%
  \textbf{\bibinfo {volume} {84}},\ \bibinfo {pages} {4693} (\bibinfo {month}
  {May}\ \bibinfo {year} {2000})%
  \bibAnnoteFile{NoStop}{Leonard00}%
\bibitem{Delerue97}%
  \BibitemOpen
  \bibfield{author}{%
  \bibinfo {author} {\bibfnamefont{C.}~\bibnamefont{Delerue}}, \bibinfo
  {author} {\bibfnamefont{M.}~\bibnamefont{Lannoo}},\ and\ \bibinfo {author}
  {\bibfnamefont{G.}~\bibnamefont{Allan}},\ }%
  \bibfield{journal}{%
  \Doi{10.1103/PhysRevB.56.15306}{\bibinfo {journal} {Phys. Rev. B}}\ }%
  \textbf{\bibinfo {volume} {56}},\ \bibinfo {pages} {15306} (\bibinfo {month}
  {Dec}\ \bibinfo {year} {1997})%
  \bibAnnoteFile{NoStop}{Delerue97}%
\bibitem{Delerue03}%
  \BibitemOpen
  \bibfield{author}{%
  \bibinfo {author} {\bibfnamefont{C.}~\bibnamefont{Delerue}}, \bibinfo
  {author} {\bibfnamefont{M.}~\bibnamefont{Lannoo}},\ and\ \bibinfo {author}
  {\bibfnamefont{G.}~\bibnamefont{Allan}},\ }%
  \bibfield{journal}{%
  \Doi{10.1103/PhysRevB.68.115411}{\bibinfo {journal} {Phys. Rev. B}}\ }%
  \textbf{\bibinfo {volume} {68}},\ \bibinfo {pages} {115411} (\bibinfo {month}
  {Sep}\ \bibinfo {year} {2003})%
  \bibAnnoteFile{NoStop}{Delerue03}%
\bibitem{Jancu98}%
  \BibitemOpen
  \bibfield{author}{%
  \bibinfo {author} {\bibfnamefont{J.-M.}\ \bibnamefont{Jancu}}, \bibinfo
  {author} {\bibfnamefont{R.}~\bibnamefont{Scholz}}, \bibinfo {author}
  {\bibfnamefont{F.}~\bibnamefont{Beltram}},\ and\ \bibinfo {author}
  {\bibfnamefont{F.}~\bibnamefont{Bassani}},\ }%
  \bibfield{journal}{%
  \Doi{10.1103/PhysRevB.57.6493}{\bibinfo {journal} {Phys. Rev. B}}\ }%
  \textbf{\bibinfo {volume} {57}},\ \bibinfo {pages} {6493} (\bibinfo {month}
  {Mar}\ \bibinfo {year} {1998})%
  \bibAnnoteFile{NoStop}{Jancu98}%
\bibitem{Bylander87}%
  \BibitemOpen
  \bibfield{author}{%
  \bibinfo {author} {\bibfnamefont{D.~M.}\ \bibnamefont{Bylander}}\ and\
  \bibinfo {author} {\bibfnamefont{L.}~\bibnamefont{Kleinman}},\ }%
  \bibfield{journal}{%
  \Doi{10.1103/PhysRevLett.59.2091}{\bibinfo {journal} {Phys. Rev. Lett.}}\ }%
  \textbf{\bibinfo {volume} {59}},\ \bibinfo {pages} {2091} (\bibinfo {month}
  {Nov}\ \bibinfo {year} {1987})%
  \bibAnnoteFile{NoStop}{Bylander87}%
\bibitem{note_offset_shift}%
  \BibitemOpen
  \bibinfo {note} {After self-consistency, the valence-band offset has a
  relatively weak dependence on $V_0$: $\Delta E_{\rm v}=0.28 - 0.13 V_{0}$
  ($V_0$ and $\Delta E_{\rm v}$ in eV).}%
  \bibAnnoteFile{Stop}{note_offset_shift}%
\bibitem{Flores87}%
  \BibitemOpen
  \bibfield{author}{%
  \bibinfo {author} {\bibfnamefont{F.}~\bibnamefont{Flores}}\ and\ \bibinfo
  {author} {\bibfnamefont{C.}~\bibnamefont{Tejedor}},\ }%
  \bibfield{journal}{%
  \bibinfo {journal} {J. Phys. C}\ }%
  \textbf{\bibinfo {volume} {20}},\ \bibinfo {pages} {145} (\bibinfo {year}
  {1987}),\ \url{http://stacks.iop.org/0022-3719/20/i=2/a=001}%
  \bibAnnoteFile{NoStop}{Flores87}%
\bibitem{Tersoff84}%
  \BibitemOpen
  \bibfield{author}{%
  \bibinfo {author} {\bibfnamefont{J.}~\bibnamefont{Tersoff}},\ }%
  \bibfield{journal}{%
  \Doi{10.1103/PhysRevLett.52.465}{\bibinfo {journal} {Phys. Rev. Lett.}}\ }%
  \textbf{\bibinfo {volume} {52}},\ \bibinfo {pages} {465} (\bibinfo {month}
  {Feb}\ \bibinfo {year} {1984})%
  \bibAnnoteFile{NoStop}{Tersoff84}%
\bibitem{Brandbyge02}%
  \BibitemOpen
  \bibfield{author}{%
  \bibinfo {author} {\bibfnamefont{M.}~\bibnamefont{Brandbyge}}, \bibinfo
  {author} {\bibfnamefont{J.-L.}\ \bibnamefont{Mozos}}, \bibinfo {author}
  {\bibfnamefont{P.}~\bibnamefont{Ordej\'on}}, \bibinfo {author}
  {\bibfnamefont{J.}~\bibnamefont{Taylor}},\ and\ \bibinfo {author}
  {\bibfnamefont{K.}~\bibnamefont{Stokbro}},\ }%
  \bibfield{journal}{%
  \Doi{10.1103/PhysRevB.65.165401}{\bibinfo {journal} {Phys. Rev. B}}\ }%
  \textbf{\bibinfo {volume} {65}},\ \bibinfo {pages} {165401} (\bibinfo {month}
  {Mar}\ \bibinfo {year} {2002})%
  \bibAnnoteFile{NoStop}{Brandbyge02}%
\bibitem{Kazymyrenko08}%
  \BibitemOpen
  \bibfield{author}{%
  \bibinfo {author} {\bibfnamefont{K.}~\bibnamefont{Kazymyrenko}}\ and\
  \bibinfo {author} {\bibfnamefont{X.}~\bibnamefont{Waintal}},\ }%
  \bibfield{journal}{%
  \Doi{10.1103/PhysRevB.77.115119}{\bibinfo {journal} {Phys. Rev. B}}\ }%
  \textbf{\bibinfo {volume} {77}},\ \bibinfo {pages} {115119} (\bibinfo {month}
  {Mar}\ \bibinfo {year} {2008})%
  \bibAnnoteFile{NoStop}{Kazymyrenko08}%
\bibitem{Soreni08}%
  \BibitemOpen
  \bibfield{author}{%
  \bibinfo {author} {\bibfnamefont{M.}~\bibnamefont{Soreni-Harari}}, \bibinfo
  {author} {\bibfnamefont{N.}~\bibnamefont{Yaacobi-Gross}}, \bibinfo {author}
  {\bibfnamefont{D.}~\bibnamefont{Steiner}}, \bibinfo {author}
  {\bibfnamefont{A.}~\bibnamefont{Aharoni}}, \bibinfo {author}
  {\bibfnamefont{U.}~\bibnamefont{Banin}}, \bibinfo {author}
  {\bibfnamefont{O.}~\bibnamefont{Millo}},\ and\ \bibinfo {author}
  {\bibfnamefont{N.}~\bibnamefont{Tessler}},\ }%
  \bibfield{journal}{%
  \Doi{10.1021/nl0732171}{\bibinfo {journal} {Nano Lett.}}\ }%
  \textbf{\bibinfo {volume} {8}},\ \bibinfo {pages} {678} (\bibinfo {year}
  {2008}),\ \url{http://pubs.acs.org/doi/abs/10.1021/nl0732171}%
  \bibAnnoteFile{NoStop}{Soreni08}%
\bibitem{Aldakov06}%
  \BibitemOpen
  \bibfield{author}{%
  \bibinfo {author} {\bibnamefont{{D. Aldakov}}}, \bibinfo {author}
  {\bibnamefont{{F. Chandezon}}}, \bibinfo {author} {\bibnamefont{{R. De
  Bettignies}}}, \bibinfo {author} {\bibnamefont{{M. Firon}}}, \bibinfo
  {author} {\bibnamefont{{P. Reiss}}},\ and\ \bibinfo {author}
  {\bibnamefont{{A. Pron}}},\ }%
  \bibfield{journal}{%
  \Doi{10.1051/epjap:2006144}{\bibinfo {journal} {Eur. Phys. J. Appl. Phys.}}\
  }%
  \textbf{\bibinfo {volume} {36}},\ \bibinfo {pages} {261} (\bibinfo {year}
  {2006}),\ \url{http://dx.doi.org/10.1051/epjap:2006144}%
  \bibAnnoteFile{NoStop}{Aldakov06}%
\bibitem{Bastide97}%
  \BibitemOpen
  \bibfield{author}{%
  \bibinfo {author} {\bibfnamefont{S.}~\bibnamefont{Bastide}}, \bibinfo
  {author} {\bibfnamefont{R.}~\bibnamefont{Butruille}}, \bibinfo {author}
  {\bibfnamefont{D.}~\bibnamefont{Cahen}}, \bibinfo {author}
  {\bibfnamefont{A.}~\bibnamefont{Dutta}}, \bibinfo {author}
  {\bibfnamefont{J.}~\bibnamefont{Libman}}, \bibinfo {author}
  {\bibfnamefont{A.}~\bibnamefont{Shanzer}}, \bibinfo {author}
  {\bibfnamefont{L.}~\bibnamefont{Sun}},\ and\ \bibinfo {author}
  {\bibfnamefont{A.}~\bibnamefont{Vilan}},\ }%
  \bibfield{journal}{%
  \Doi{10.1021/jp9626935}{\bibinfo {journal} {J. Phys. Chem. B}}\ }%
  \textbf{\bibinfo {volume} {101}},\ \bibinfo {pages} {2678} (\bibinfo {year}
  {1997})%
  \bibAnnoteFile{NoStop}{Bastide97}%
\bibitem{Ashkenasy02}%
  \BibitemOpen
  \bibfield{author}{%
  \bibinfo {author} {\bibfnamefont{G.}~\bibnamefont{Ashkenasy}}, \bibinfo
  {author} {\bibfnamefont{D.}~\bibnamefont{Cahen}}, \bibinfo {author}
  {\bibfnamefont{R.}~\bibnamefont{Cohen}}, \bibinfo {author}
  {\bibfnamefont{A.}~\bibnamefont{Shanzer}},\ and\ \bibinfo {author}
  {\bibfnamefont{A.}~\bibnamefont{Vilan}},\ }%
  \bibfield{journal}{%
  \Doi{10.1021/ar990047t}{\bibinfo {journal} {Accounts Chem. Res.}}\ }%
  \textbf{\bibinfo {volume} {35}},\ \bibinfo {pages} {121} (\bibinfo {month}
  {FEB}\ \bibinfo {year} {2002})%
  \bibAnnoteFile{NoStop}{Ashkenasy02}%
\bibitem{Bauer83}%
  \BibitemOpen
  \bibfield{author}{%
  \bibinfo {author} {\bibfnamefont{R.~S.}\ \bibnamefont{Bauer}}, \bibinfo
  {author} {\bibfnamefont{P.}~\bibnamefont{Zurcher}},\ and\ \bibinfo {author}
  {\bibfnamefont{J.}~\bibnamefont{Henry W.~Sang}},\ }%
  \bibfield{journal}{%
  \bibinfo {journal} {Applied Physics Letters}\ }%
  \textbf{\bibinfo {volume} {43}},\ \bibinfo {pages} {663} (\bibinfo {year}
  {1983})%
  \bibAnnoteFile{NoStop}{Bauer83}%
\bibitem{Algra11}%
  \BibitemOpen
  \bibfield{author}{%
  \bibinfo {author} {\bibfnamefont{R.~E.}\ \bibnamefont{Algra}}, \bibinfo
  {author} {\bibfnamefont{M.}~\bibnamefont{Hocevar}}, \bibinfo {author}
  {\bibfnamefont{M.~A.}\ \bibnamefont{Verheijen}}, \bibinfo {author}
  {\bibfnamefont{I.}~\bibnamefont{Zardo}}, \bibinfo {author}
  {\bibfnamefont{G.~G.~W.}\ \bibnamefont{Immink}}, \bibinfo {author}
  {\bibfnamefont{W.~J.~P.}\ \bibnamefont{van Enckevort}}, \bibinfo {author}
  {\bibfnamefont{G.}~\bibnamefont{Abstreiter}}, \bibinfo {author}
  {\bibfnamefont{L.~P.}\ \bibnamefont{Kouwenhoven}}, \bibinfo {author}
  {\bibfnamefont{E.}~\bibnamefont{Vlieg}},\ and\ \bibinfo {author}
  {\bibfnamefont{E.~P. A.~M.}\ \bibnamefont{Bakkers}},\ }%
  \bibfield{journal}{%
  \bibinfo {journal} {Nano Letters}\ }%
  \textbf{\bibinfo {volume} {11}},\ \bibinfo {pages} {1690} (\bibinfo {year}
  {2011})%
  \bibAnnoteFile{NoStop}{Algra11}%
\bibitem{Xuan2010}%
  \BibitemOpen
  \bibfield{author}{%
  \bibinfo {author} {\bibfnamefont{X.~P.~A.}\ \bibnamefont{Gao}}, \bibinfo
  {author} {\bibfnamefont{G.}~\bibnamefont{Zheng}},\ and\ \bibinfo {author}
  {\bibfnamefont{C.~M.}\ \bibnamefont{Lieber}},\ }%
  \bibfield{journal}{%
  \bibinfo {journal} {Nano Letters}\ }%
  \textbf{\bibinfo {volume} {10}},\ \bibinfo {pages} {547} (\bibinfo {year}
  {2010})%
  \bibAnnoteFile{NoStop}{Xuan2010}%
\end{thebibliography}%

\end{document}